\documentclass[%
 reprint,
superscriptaddress,
 amsmath,amssymb,mathtools,
pre,
floatfix,
]{revtex4-2}

\usepackage{graphicx}% Include figure files
\usepackage{dcolumn}% Align table columns on decimal point
\usepackage{bm}% bold math
\usepackage{hyperref}% add hypertext capabilities
\usepackage{mathtools}
\usepackage{tikz}
\usepackage{physics}
\usepackage{placeins}
\usepackage{dsfont}
\usepackage{ifthen}
\usepackage{enumitem} 

\newcommand{\resubmit}[1]{{#1}}

\newcommand{\hbarE}{{\hbar_{\text{eff}}}}

\newcommand{\tLoc}{\ensuremath{\tau_{\rm s}}}
\newcommand{\tLeak}{\ensuremath{\tau_{\rm L}}}
\newcommand{\tEhr}{\ensuremath{\tau_{\rm E}}}

\newcommand{\ha}[1]{\ensuremath{\hat{a}_{#1}}}
\newcommand{\had}[1]{\ensuremath{\hat{a}^{\dagger}_{#1}}}

\newcommand{\ls}{\ensuremath{\lambda_{\rm s}}}

\renewcommand{\lq}{\ensuremath{\lambda_{\rm q}}}
\renewcommand{\d}{\ensuremath{\operatorname{d}\!}}

\setcitestyle{numbers,square,sort&compress}

\begin{document}

\preprint{PRE }

\title{Dynamical transition from localized to uniform scrambling in locally hyperbolic systems}% Force line breaks with \\

\author{Mathias Steinhuber} %\email{mathias.steinhuber@ur.de}
\affiliation{%
 Institut f\"ur  Theoretische Physik, Universit\"at Regensburg, 
  93040 Regensburg, 
 Germany
 }
\author{Peter Schlagheck} %\email{peter.schlagheck@uliege.be}
\affiliation{%
 CESAM Research Unit, University of Liege, 4000 Liège, Belgium
}

\author{Juan Diego Urbina}%
 \affiliation{%
 Institut f\"ur  Theoretische Physik, Universit\"at Regensburg, 
 93040 Regensburg, 
 Germany
 }
\author{Klaus Richter}%
\affiliation{%
 Institut f\"ur  Theoretische Physik, Universit\"at Regensburg,
  93040 Regensburg, 
 Germany
}

\date{\today}% It is always \today, today,

\begin{abstract}
Fast scrambling of quantum correlations, reflected by the exponential growth of Out-of-Time-Order Correlators (OTOCs) on short pre-Ehrenfest time scales, is commonly considered as a major quantum signature of unstable dynamics in quantum systems with a classical limit. In two recent works \resubmit{[Phys. Rev. Lett. \textbf{123}, 160401 (2019)]} %by Hummel et al.~\cite{Hummel2019} and by Xu et al.~\cite{Scaffidi2020}
and \resubmit{[Phys. Rev. Lett. \textbf{124}, 140602 (2020)]}, a significant difference in the  scrambling rate of integrable (many-body) systems was observed, depending on the initial state being semiclassically localized around unstable fixed points or fully delocalized (infinite temperature). Specifically, the quantum Lyapunov exponent $\lambda_{\rm q}$ quantifying the OTOC growth is given, respectively, by $\lambda_{\rm q}=2\lambda_{\rm s}$ or $\lambda_{\rm q}=\lambda_{\rm s}$ in terms of the stability exponent $\lambda_{\rm s}$ of the hyperbolic fixed point. Here we show that a wave packet, initially localized around this fixed point, features a distinct {\it dynamical} transition between these two regions. We present an analytical semiclassical approach providing a physical picture of this phenomenon and support our findings by extensive numerical simulations in the whole parameter range of locally unstable dynamics of a Bose-Hubbard dimer. %Our results suggest that the existence of this transition is a hallmark of unstable separatrix dynamics in integrable systems. 
\resubmit{Our results suggest that the existence of this crossover is a hallmark of unstable separatrix dynamics in integrable systems, thus opening the possibility to distinguish the latter, on the basis of this particular observable, from genuine chaotic dynamics generally featuring uniform exponential growth of the OTOC.}
\end{abstract}
\maketitle
%\keywords{Integrable Dynamics, Separatrix Motion, Out-of-Time-Order Correlator, Scrambling}

%%%%%%%%%%%%%%%% intro %%%%%%%%%%%%%%%%%%%%

\section{Introduction}
The scrambling of quantum correlations is an ubiquitous phenomenon  across the physics of interacting many-body systems \cite{SwingleTutorial,Richter2022}. Its connection to quantum chaos  has been established in systems ranging from models for black holes  \cite{Maldacena_2016,Hayden_2007,Sekino:2008he,Kitaev2019} to realistic many-body systems such as the SYK-model \cite{standfordOTOC2021}, even comprising systems without a classical limit \cite{OTOCfermions}.
Due to the appealing connection with the powerful concepts of quantum chaos, Out-of-Time-Ordered Correlators (OTOCs) \cite{classic} 
% with their short-time exponential growth 
represent a major probe of scrambling \resubmit{(see, however, \footnote{\resubmit{Although the exponential behavior of OTOCs is a hallmark of classical dynamical stability and therefore present in single- and few-body systems as well, its use as a measure of the scrambling of correlations is appropriate only in the many-body context \cite{SwingleTutorial,Maldacena_2016}}}}) and thus have received a swiftly increase of theoretical interest \cite{Maldacena_2016,SwingleTutorial} that has driven efforts for experimental proposals \cite{proposalExperimentOTOC} and realizations \cite{Garttner2017,NMR_OTOC}.

In systems with a semiclassical regime, fast scrambling is considered an unambiguous indicator of classical (mean-field) instabilities \cite{Richter2022}. As such, Bose-Hubbard systems, with their well understood and controlled classical (mean-field) limit and a large semiclassical region of state space, are prime models to study imprints of scrambling~\cite{dimerBH,Shen2017,Bohrdt2017,Josef2018}.
%\cite{Maldacena_2016,SwingleTutorial, Hummel2019, argentinians} 
% transition to the actual topic
Recently it was shown~\cite{Hummel2019,Papparlardi2018} that an initial exponential growth of OTOCs does not necessarily imply chaotic dynamics of the system's classical counterpart, {\em i.e.} such OTOC behavior alone cannot serve as clear-cut probe of quantum chaos. These works~\cite{Hummel2019,Papparlardi2018} and further ones picking up the same idea \cite{Scaffidi2020,Geiger2021,Santos1,Santos2} show that for quantum (many-body) systems with a classical limit and a semiclassical regime it is sufficient to have local instabilities in a (possibly integrable) phase space to generate exponentially growing OTOCs. Several examples of this situation have been \resubmit{numerically} studied \cite{Pilatowsky2020OTOCregularSystem,Kidd2021Dicke}, including basic models such as an inverted harmonic oscillator \cite{jhep11(2020)068}.

\resubmit{It is worth mentioning that, with respect to their classical limit, many-body quantum systems fall into one of two broad classes depending on the dimension of the (local) Hilbert space where the single-particle dynamics takes place. On one extreme, we have systems with small local Hilbert space like spin $1/2$ chains or fermionic fields in a lattice. In this case a classical limit can be defined, but the system lacks a notion of classical regime. On the other extreme of large local Hilbert space we have a precise classical limit, together with a precise notion of when and how the it actually reproduces faithfully the quantum dynamics \cite{Richter2022}. Systems described by interacting bosonic fields in the lattice are a paradigmatic example of the later. Their classical regime is defined by large number occupations instead of the large energy condition typical of the single-particle systems  which we utilize in the case of the Bose Hubbard dimer (see \footnote{ \resubmit{
Interestingly, under specific circumstances, paradigmatic examples of systems with small local Hilbert space do have a well defined classical regime. This is the case with, for example, spin $1/2$ chains with all-to-all interactions \cite{Papparlardi2018}.}
}).}

Generically, the prime example of the mechanism for an exponential OTOC growth in an integrable system is the existence of an unstable (hyperbolic) fixed point. Although, by definition, all Lyapunov exponents $\lambda_{\rm L}$ are zero in integrable systems, the classical dynamics around fixed points is locally hyperbolic if they have at least one positive stability exponent $\ls>0$. This type of instability will be considered here.
In the early time regime, defined up to a time scale depending logarithmically on the effective Planck constant $\hbarE$, the OTOCs involving dynamics around unstable points of many-body integrable systems display two markedly different behaviors. In Refs.~\cite{Hummel2019} and \cite{Scaffidi2020}, the quantum Lyapunov exponents $\lq$ quantifying the OTOC growth rate are compared to the stability exponents of dominant unstable fixed points of the corresponding classical mean-field dynamics yielding good agreement with~$2\ls$ \cite{Hummel2019} or $\ls$ \cite{Scaffidi2020}, respectively.

In this paper we resolve this apparent discrepancy with regard to the operator growth and provide a unified dynamical mechanism explaining the two results in a comprehensive way. We demonstrate that there is a universal~$2\ls$ to $\ls$ transition for the OTOC growth rate $\lq$  for dynamics around unstable fixed points in the pre-Ehrenfest time regime which interpolates between these two limits.
Furthermore, the crossover develops a kink in the strictly classical limit $\hbarE \to 0$. 
Moreover, we show that this $2\ls$ to $\ls$ transition is related to an underlying dynamical \resubmit{crossover} of the initial quantum state in a phase-space representation and argue that the $2\ls$ to $\ls$ transition is a hallmark of integrable systems. 
We propose that this effect can thus be used to distinguish chaotic and integrable systems by properly analyzing the growth behavior of OTOCs at pre-Ehrenfest time scales.

% contents
The paper is structured into three further sections. In Sec.~\ref{sec:OTOC_theory}, we present a heuristic argument for OTOCs to exhibit different exponential regimes and the corresponding $2\ls$ to~$\ls$ \resubmit{crossover}. 
In order to verify the picture put forward, we perform an extensive study of the Bose-Hubbard dimer in Sec.~\ref{sec:boseHubbard}. There, we start with the definition of the dimer and a study of the classical mean-field system including an analytical study of the classical OTOC. 
Then, numerical results for the OTOCs follow obtained from extensive simulations that display excellent agreement with the classical results for the OTOCs. Furthermore, we show how one can tune the~$2\ls$ to~$\ls$-transition, and study its robustness with regard to changing the system parameters. 
In Sec.~\ref{sec:conclusion} we summarize our findings and discuss their possible extension to non-integrable systems.

%%%%%%%%%%%%%%%%%%% end intro %%%%%%%%%%%%%%%%%%%%%
\section{Out-of-Time-Order Correlator in~integrable~systems with local hyperbolicity}%
\label{sec:OTOC_theory}%
Our goal in this section is to refine the pre-Ehrenfest theory for scrambling around hyperbolic fixed points in integrable systems. In particular we attempt to relax the localization properties of the initial state considered in~\cite{Hummel2019,Scaffidi2020}.
The OTOC for two operators $\hat{A}$, $\hat{B}$ with respect to a state $\hat{\rho}$ is defined by 
\begin{align}
    \mathbf{ C}(t) \,=\, \operatorname{tr} \big\{\hat{\rho} \big| [\hat{A}(t),\hat{B} ] \big|^2 \big\},
    \label{eq:OTOC_def} 
\end{align}
which is by itself a modulus-squared commutator. When this squared commutator is expanded in individual correlators one obtains, besides contributions that admit a standard time ordering, extra irreducibly un-ordered correlations \cite{Maldacena_2016}, with anomalous dynamical behavior that are the central object of study.  

The long-time~(post-Ehrenfest) saturation of generic OTOCs has been subject of several studies, both in the chaotic \cite{Josef2018,argentinians} and integrable \cite{Hummel2019,Fortes} regimes where interference effects beyond a pure quasiclassical (Truncated-Wigner like-) approach appear~\cite{Polkovnikov2011}.

Here, however, our focus is the short time scales, where a quasiclassical approach based on the Wigner-Moyal expansion, which is a regular expansion around $\hbarE=0$ \cite{schleich01,Kim1991,connell2008}, is perfectly appropriate. Keeping only leading-order terms in $\hbarE$, one obtains %(see App.~\ref{sec:appendixWignerMoyal})\rc{maybe an appendix? YES but not in present arXiv version, just put as reference the quantum optics in phase space book}

    \begin{align}
        \begin{split}
            \mathbf{ C}(t) \,=\,& \hbarE^{2} \langle W_{\rho}(\Vec{q}_{0},\Vec{p}_{0}) \big|\big\{ A_{\rm W}(\Vec{q}_{0},\Vec{p}_{0},t), B_{\rm W}(\Vec{q}_{0},\Vec{p}_{0}) \big\}  \big|^{2}\rangle_{\text{PS}}
            \\
            &+ O(\hbarE^{3}),
        \end{split}
         \label{eq:WignerWeyl}
    \end{align}
    where $A_{W}$, $B_{W}$ are the Wigner-transforms of the operators $\hat{A},\hat{B}$ and $W_{\rho}(\Vec{q}_{0},\Vec{p}_{0})$   is the Wigner-distribution corresponding to the state $\hat{\rho}$.  A detailed derivation of Eq.~\eqref{eq:WignerWeyl} is in App.~\ref{sec:appendixWignerMoyal}.

    Further, $\langle  . \rangle_{\text{PS}}$ indicates integration over the whole classical (mean-field) phase space parametrized by the canonical pairs $(\Vec{q}_{0},\Vec{p}_{0})$. The Heisenberg time dynamics of the quantum operator is mapped to time dynamics of the classical observable $A(q_{0},p_{0},t) =A(q(q_{0},p_{0},t),p(q_{0},p_{0},t)) $ which arises from the classical propagation of the initial condition $(q_{0},p_{0})$.
    
    The effective Plank constant $\hbarE$ has different expressions in different contexts. It is given by the usual Plank constant divided a typical action $\hbar/S_{\rm typ}$ in single particle cases \cite{Gutzwiller1991,Haake2001,brack1997}, and the inverse of the total spin quantum number $1/S$ for spin systems \cite{Waltner2017PRL,Waltner2020PRE}. In the case of interest here, interacting bosonic systems, $\hbarE = \frac{1}{N}$  is given by the inverse of the total particle number $N$ after convenient rescaling of the interaction strength \cite{Richter2022,Engl2015PRE}.
    
    Without loss of generality, we choose the operators $\hat{A}=\hat{q}$ and $\hat{B}=\hat{p}$, as they are hermitian and their classical counterparts are generalized coordinates or momenta. Therefore, at leading order in the Wigner-Moyal expansion, Eq.~\eqref{eq:WignerWeyl},
    we drop the index ${}_W$ and take the pure classical phase space functions. This choice of the operators simplifies the classical Poisson-brackets, that are now given by an element of the stability matrix $\frac{\partial \vec{x}(t)}{\partial \vec{x}(0)}$ with $\vec{x} = (\vec{q},\vec{p})$ \cite{tabor1989chaos,wiggins2003}. 

    The classical limit of the OTOC in Eq.~(\ref{eq:WignerWeyl}) is so far a completely general result. Under the assumption of local instability, however, the leading order of exponential growth is given by the maximal local exponent $\lambda(\vec{q}_{0},\vec{p}_{0})$ of the stability matrix:
    \begin{align}
         \mathbf{ C}(t) ~\sim~& \hbarE^{2} \langle W_{\rho}(\Vec{q}_{0},\Vec{p}_{0}) \exp\big\{ 2 \lambda(\Vec{q}_{0},\Vec{p}_{0}) t\big\}  \rangle_{\text{PS}}.
         \label{eq:WignerWeylFP}
    \end{align}
    
    At this point, it is convenient to introduce two local time-scales:
    \begin{itemize}
        \item After the local ergodic time $\tLoc= 1/\lambda (\Vec{q}_{0},\Vec{p}_{0})$,  the exponential growth of the OTOC begins to be visible. Before $\tLoc$, we have sub-exponential/polynomial behavior linked to system-specific mechanisms.
        \item The Wigner-Weyl approximation breaks down when the leading order of the integrand in Eq.~\eqref{eq:WignerWeyl} becomes large compared to $\hbarE^{2}$. This breakdown defines the local Ehrenfest time $\tEhr= \lambda^{-1}(\Vec{q}_{0},\Vec{p}_{0})  \log (1/\hbarE)$ and in our many-body case $\tEhr= \lambda^{-1}(\Vec{q}_{0},\Vec{p}_{0}) \log N$.
    \end{itemize}
    
    We exploit the experimentally tunable localization features of quantum mechanical states \cite{squeezingBEC2008} as a tool to probe the local unstable dynamics around a hyperbolic fixed point (FP)  and consider a coherent-like state~$\hat{\rho}$  centered around it. 
    In the linearized region around the FP the dynamics can be precisely described, and we can express $\lambda(\vec{q}_{0},\vec{p}_{0})$ by the maximal stability exponent $\ls$ of the FP. 
    In general, however, the linearized region is bounded and we express this by the fact that the relation $\{ A(\vec{q}_{0},\vec{p}_{0},t), B(\vec{q}_{0},\vec{p}_{0})\} = e^{\ls t} $ is valid only if the unstable manifold coordinate/projection $u(\vec{q}_{0}$, $\vec{p}_{0})$ is smaller than a threshold $c>0$. 
    For times larger than zero, the exponential growth $u(\vec{q}_{0},\vec{p}_{0},t)= u(\vec{q}_{0},\vec{p}_{0}) e^{\ls t}$ is only valid if the linearized region $u(\vec{q}(t),\vec{p}(t),t)< c$ is still fulfilled. Afterwards we need to replace the time dynamics of the unstable manifold by a sub-exponential function. We refer to this mechanism as a kind of {\it leaking} from the linearized region \cite{Scaffidi2020}.%\rc{(cite Scaffidi)}
    
    The key observation is that, although the sub-exponential function describes the classical evolution outside and it is therefore negligibly  small compared to the exponential growth in the linearized region, its contribution is weighted by a portion of phase space that grows exponentially. These considerations allow us to heuristically account for the leaking mechanism, by modifying the Eq. \eqref{eq:WignerWeylFP} as
    \begin{align}
         \mathbf{ C}(t) ~\sim~& \hbarE^{2} e^{ 2 \ls t} \langle W_{\rho}(\Vec{q}_{0},\Vec{p}_{0}) \Theta ( c - e^{\ls t} u(\vec{q}_{0}, \vec{p}_{0}))\rangle_{\text{LR}},
         \label{eq:WignerWeylFPLeaking}
    \end{align}
    where we restrict the phase-space integration to the dominant linearized region (LR) around the FP.

\begin{figure}[h!]
        \centering
        \begin{tikzpicture}[scale=2.2]
            % Draw axes
            \draw [<->,thick] (0,2) node (yaxis) [above,rotate=90,xshift=-0.5cm] {ln $\mathbf{C}(t)$}
                |- (3,0) node (xaxis) [below] {time $t$};
            \draw[dashed] (0.5,2) -- (0.5,0) node[below] {$\tau_{s}$};
            \draw[dashed] (1.3,2) -- (1.3,0)  node[below] {$\tLeak$};
            \draw[dashed] (2.3,2) -- (2.3,0) node[below] {$\tEhr$};
            \draw[very thick] (0.5,0.5) -- (1.3,1.3)node [midway, above,yshift=0.5cm] {$\sim e^{2\ls t}$} -- (2.3,1.8) node [midway, below,yshift=-0.2cm] {$\sim e^{\ls t}$};
            \draw[very thick, dotted] (0.5,0.5) arc[start angle=140, end angle=170,radius=1cm] ;
            %\draw[very thick, dotted] (2.3,1.8) -- (2.9,1.8) node[midway,above] {\tiny chaotic};
            \draw[very thick, dotted] (2.3,1.8) cos (2.7,1.4);
            \draw  (2.45,1.35) node[above,rotate=-55.5] {\tiny integrable};
        \end{tikzpicture}        
        \caption{Expected behavior of an OTOC centered at a FP if~$\tLoc<\tLeak< \tEhr$: OTOCs grow polynomial for times shorter $\tLoc$, exponential with $2\ls$ for times shorter $\tLeak$, exponential with $\ls $ for times shorter $\tEhr$, but greater than $\tLeak$. Post-Ehrenfest time scales display oscillatory behaviour if the system is integrable~\cite{Fortes} and saturation if the system is chaotic \cite{Josef2018,PhysRevE.98.062218}. }
        \label{fig:SchematicOTOC_FP}
    \end{figure}
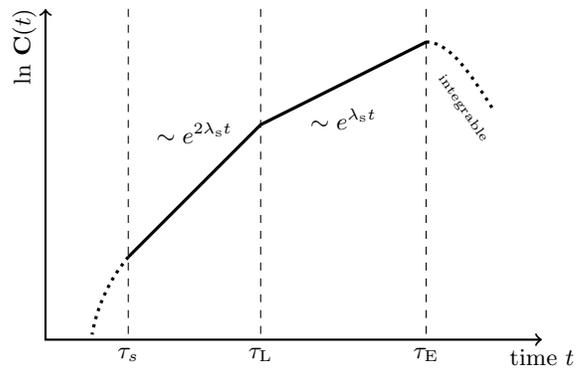

    For definiteness, let us consider now a Gaussian wave-packet with an initial linear width  $\Delta u$ along the unstable manifold $u(\vec{q},\vec{p})$.
    %Leaking affects the dynamics of the OTOC \ref{eq:WignerWeyl} for a wave-packet with an initial linear width $\Delta u$ along the unstable manifold $u(\vec{q},\vec{p})$.
    In this situation, the wave-packet reaches the boundary of the  linear region by a finite time $\tLeak = \ls^{-1} \ln \big( c/\Delta u \big)$, which we correspondingly call the leaking time. 
    After $\tLeak$, we must take the leaking of the wave-packet into account, i.e., the phase-space volume causing the exponential growth shrinks exponentially with $e^{-\ls t}$, i.e.,
    \begin{align*}
        \langle W_{\rho}(\Vec{q}_{0},\Vec{p}_{0}) \Theta ( c - e^{\ls t} u(\vec{q}_{0}, \vec{p}_{0}))\rangle_{\text{LR}} \,\sim \,
        \left\{\begin{array}{ll}
            \text{const.} &,t<\tLeak \\
            e^{-\ls t} &,t>\tLeak \\
             \end{array}\right. \! .
    \end{align*}
    Hence, the exponential growth of the OTOC in Eq. \eqref{eq:WignerWeylFPLeaking} decreases to $e^{\ls t}$. 
    This finally leads to a short-time behavior of the OTOC around an unstable fixed point given by
    \begin{align}
        \mathbf{ C }(t)~\sim~
            \left\{\begin{array}{ll}
            \text{poly.} & ,t<\tLoc \\
             e^{2\ls t}& ,\tLoc<t<\tLeak\\
             e^{\ls t} & ,\tLeak<t<\tEhr\\
             \text{osc.} & ,\tEhr < t 
             \end{array}\right.
             ,
             \label{eq:OverviewFP}
    \end{align}
     that is schematically displayed in Fig. \ref{fig:SchematicOTOC_FP} showing two exponential regions. 
    
    It is important to note that there is a hierarchy of time scales: the leaking time is only relevant if~$\tLeak < \tEhr$, otherwise the Wigner-Weyl approximation (in leading order), see Eq.~\eqref{eq:WignerWeyl}, is already invalid. 

    The initial linear width $\Delta u$ scales with some power $\hbar_{\rm eff}^{\alpha}$ for typical states. This gives an asymptotic expression for the leaking time by $\tLeak\sim \frac{\alpha}{ \ls} \log N + O(\log(c))$. Hence, we have a direct proportional relation to the Ehrenfest time $ \tLeak \sim  \alpha\tEhr$ for $\hbarE\to 0$.
    
    We see then, that one can clearly distinguish three parametric regions: $\tLeak < \tLoc$,  $\tLoc<\tLeak < \tEhr$ and $\tEhr \leq \tLeak$ (which is equivalent to $\alpha\approx 0,<1,\approx 1$):
    \begin{enumerate}[label=\roman*)]
        \item Delocalized/uniform states: $\tLeak < \tLoc$, $\alpha\approx 0$\\
        Under the assumption that there is only one  unstable FP of the classical dynamics, the OTOC is still governed by Eq.~\eqref{eq:OverviewFP} and the $e^{2\ls t}$-regime vanishes. 
        A    typical examples here are high temperature states ($T\to \infty$).
        \item Localized states:  $\tLoc<\tLeak < \tEhr$, $0< \alpha <1$\\
            In this case we have the $2\ls -\ls$ transition and asymptotically (for $N\to \infty$) we expect a sharp kink to appear at $\tLeak$.
           The prime example are the coherent states centered at a FP. They usually have a linear size of $\hbarE^{1/2}$ in all phase space directions. 
        \item Well-localized states: $\tLeak\approx \tEhr$, $\alpha \approx 1$\\
            The second $e^{\ls t}$-region is vanishing, only the one $e^{2\ls t}$-region is visible. 
            Fock states are candidates for the third class. Their linear width is $\hbarE$ in the classical occupation numbers, such that $\alpha \approx 1$ if the unstable manifold is aligned in the parallel direction.  
    \end{enumerate}
The case $\alpha>1$ is unphysical and can be excluded. The uncertainty principle requires the product of the width in all directions to be $\geq \hbarE /2$. Hence if $\alpha>1$, one direction must increase if $\hbarE\to 0$, i.e., this direction becomes delocalized if we approach the classical limit contradicting that the state is associated by a well-defined point in phase space. %Delocalization contradicts such a classical limit.% \rc{maybe explain why? I mean... a Fock state is delocalized and admits a classical limit in the sense of an ensemble...}.

At this point, we can explain the dynamical behavior of OTOCs reported in \cite{Hummel2019} and \cite{Scaffidi2020}. In the first paper, the authors investigated a number-projected coherent state which is simultaneously a Fock state. Its unstable manifold is parallel to the occupation direction \cite{thesisBenni}, therefore it falls into the case  iii) and they see only the $2\ls$-exponential window. Correspondingly, the authors of the second paper  use the infinite temperature state, hence their state directly falls into the first case and the only exponential window is given by $e^{\ls t}$.

In the next section, we numerically explore the validity Eq. \eqref{eq:OverviewFP} for a  Bose-Hubbard dimer, with the aim of carefully investigate the new case  ii), where the hierarchy of time scales $\tLoc<\tLeak<\tEhr$ implies, from our analysis, the presence of a $2\ls-\ls$ transition.

\section{Bose-Hubbard Dimer}
\label{sec:boseHubbard} 
The Bose-Hubbard dimer describes bosonic degrees of freedom occupying %on 
two discrete levels or sites. Prime physical setups are individual Josephson-junctions~\cite{superconductorJosephson2001} or cold atoms within a small two-sited optical lattice~\cite{Albiez2005,F_lling_2007,Witthaut2008,cheinet2008,Kierig2008,Tomkovi2017}. In all these cases, one ends with an effective description in terms of the following Hamiltonian
\begin{align}
    \label{eq:DefQuantumDimer}
\hat{H} = -2 J \big( \had{2}\ha{1}+\ha{2}\had{1} \big) + \frac{g}{2} \big( \had{1}{}^{2} \ha{1}^{2}  +\had{2}{}^{2} \ha{2}^{2}   \big),
\end{align}
where the parameter $J$ is the hopping and $g$ is the (local) interaction strength between particles given in units of energy. Our Hamiltonian differs from the usual dimer Hamiltonian, the hopping coefficient $2J$ (instead of $J$) is motivated to be consistent with a ring topology for higher number of wells. Consequently, the two-site ring has a doubly counted hopping term. We also introduce a new dimensionless parameter~$\Theta$ such that we have $J=\epsilon_{0}\cos \Theta$ and $g = \epsilon_{0}\frac{2}{N} \sin \Theta$, %Using $\Theta$ fixes the scale of the parameters to 
with $\epsilon_{0} = \sqrt{J^{2} + \big(g N/2\big)^{2}}$ %in units of an 
representing a global energy scale. %$\epsilon_{0}$.
We set $\epsilon_{0}=1$ for a convenient unit system, yielding also the time unit $\hbar/\epsilon_{0}=1$.
The parameter space is thus compactified to $\Theta \in [-\pi/2,\pi/2]$.
%Furthermore, this parametrization of $J$ and $g$ makes the spectrum linearly in the particle number $N$ in leading order. %\rc{Sorry man, you must introduce here physical units, as now J and g appear dimensionless}. 

\subsection{Classical mean-field limit}
We follow the standard approach \cite{negele1995quantum} to derive the classical limit for bosons and replace the operators by complex numbers 
\begin{align*}
    \ha{j}, \had{j} ~\longmapsto&~ \psi_{j},\psi^{\ast}_{j}
\end{align*}
within the normal-ordered quantum Hamiltonian in Eq.~\eqref{eq:DefQuantumDimer} to obtain a classical mean-field system. 
%Hamilton's equations of motion 
The discrete nonlinear Schr\"odinger equation $i \dot{\psi}_{j} = \frac{\partial H}{\partial \psi_{j}^{\ast}}$ yields Hamilton's equations of motion that define the classical dynamics.
Due to the conserved total particle number $N$, we define new set of conjugated classical variables
\begin{align*}
    \begin{matrix*}[l]
        N  =  n_1 + n_2 \,, &\phi  =  \frac{1}{2}( \varphi_1 + \varphi_2 ) \,, \\
        n  =  \frac{1}{2}( n_1 - n_2) \,, &\varphi  =  \varphi_1 - \varphi_2 - \pi \,,
    \end{matrix*}
\end{align*}
where the two mean fields $\psi_{j}=\sqrt{n_{j}}e^{i\varphi_{j}}$ are written in phase~$\varphi_{j}$ and occupations~$n_{j}$. Hence, the Hamiltonian takes the form
\begin{eqnarray}
        H(N,\phi, n,\varphi) & = & 2\cos{\Theta} \sqrt{N^2 - 4 n^2} \cos \varphi \nonumber \\ && +\sin\Theta \Big( \frac{2n^2 }{N} +  \frac{N}{2} \Big)\,.    \label{eq:classicalHamiltonian}
\end{eqnarray}
We can reduce the dynamics to a one-dimensional system with a single conjugated pair $(z=2n/N,\varphi= \varphi_{1}-\varphi_{2}-\pi)$ given by the population inversion and relative phase~\cite{Campbell2020Dimer}. 
With these coordinates, we can reduce the equations of motion to two coupled real-valued ODEs
\begin{align}
    \begin{split}
        \dot{z} &= -4\cos\Theta \sqrt{1-z^{2}} \sin \varphi,
        \\
        \dot{\varphi} &=  4\cos\Theta \frac{z\cos\varphi}{\sqrt{1-z^{2}}} - 2\sin \Theta \, z,
    \end{split}
    \label{eq:EquationOfMotion}
\end{align}
%{\color{red} forgot hopping $J$ I think, multiply everything by $J= \cos(\Theta)$, to check!}
which is a one-degree-of-freedom system. 
Conveniently, the mean-field system deriving from Eq.~\eqref{eq:EquationOfMotion} is 
integrable and can thus be exactly solved up to quadrature.
%exactly solvable and  independent of the particle number $N$. 

\subsection{Fixed points}

A straightforward calculation shows that there are two fixed points (FPs), at ($z=0$, $\varphi=\pi$) and ($z=0$, $\varphi=0$), which are independent of the system parameter $\Theta$. 
We call these two FPs the \textit{in-phase} and \textit{off-phase} FP, since both have homogeneous occupations and a zero or $\pi$ phase difference between site 1 and 2. We set the zero point for the relative phase $\varphi$ to the unstable off-phase FP. 

\begin{figure}[t]
    \centering
    \includegraphics[width=1\linewidth]{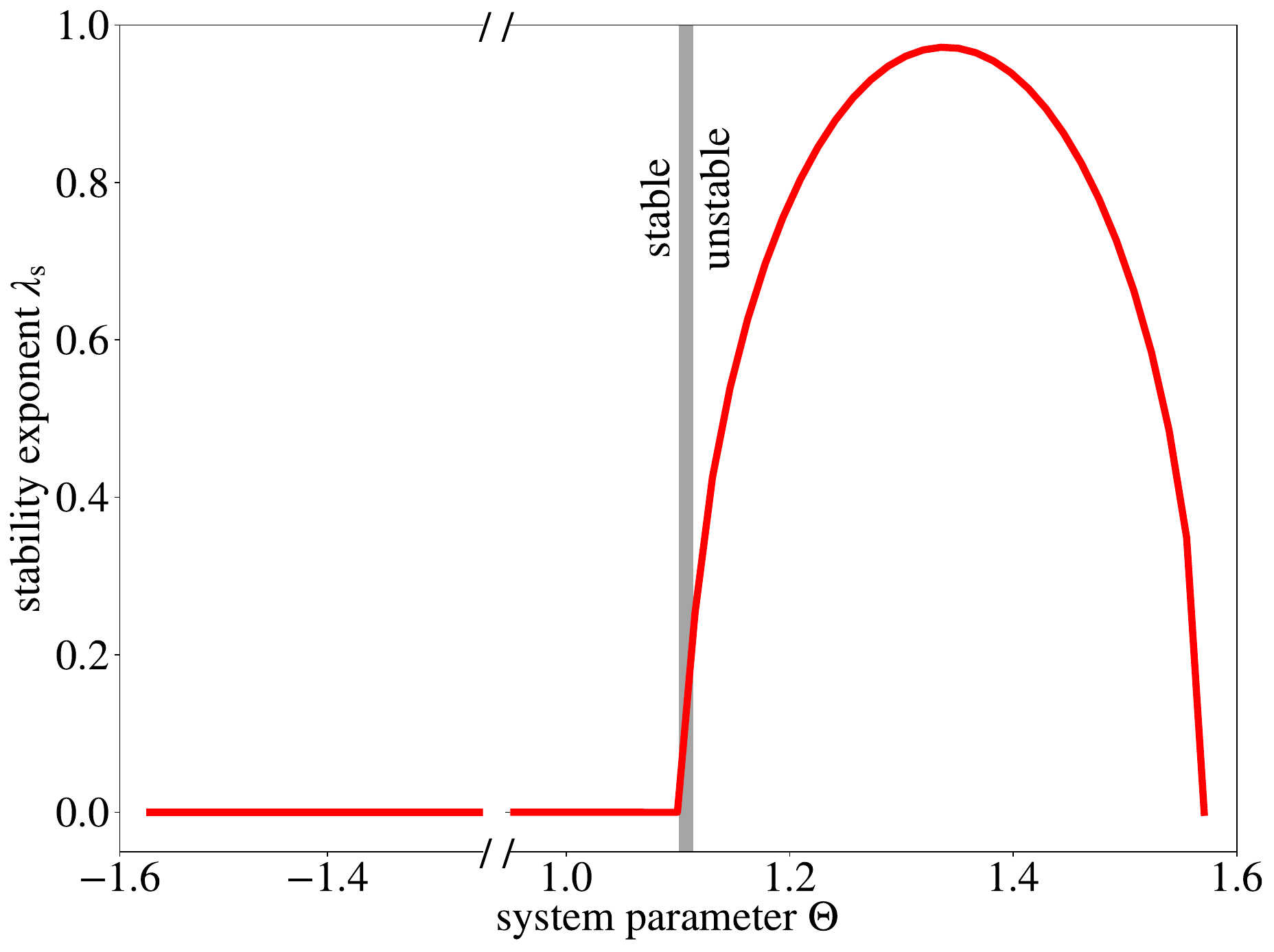}
    \caption{Stability exponent of the off-phase FP over the whole parameter space. A change from stable to unstable behaviour occurs at the bifurcation point $\Theta=\arctan 2$.}
    \label{fig:StablHomFP}
\end{figure}

Two bifurcations appear: at $\Theta = - \arctan 2$ for the in-phase FP and at $\Theta =  \arctan 2$ for off-phase FP. 
We restrict our discussion to the off-phase FP, since there is a symmetry between these two PFs under a change of sign of the parameter~$\Theta$.
%%%%%%%%%%%%5
 % need to check if stability analysis in the reduced form give the same diagram!!!!!
%%%%%%%%
The stability diagram of the off-phase FP, displayed in Fig.~\ref{fig:StablHomFP}, shows the bifurcation at $\arctan 2$, where the stability exponent becomes positive. Its maximum $\ls=0.97$ is reached at $\Theta_{\ast}\approx1.35$.

\begin{figure}[h!]
    \centering
    \includegraphics[width=1\linewidth]{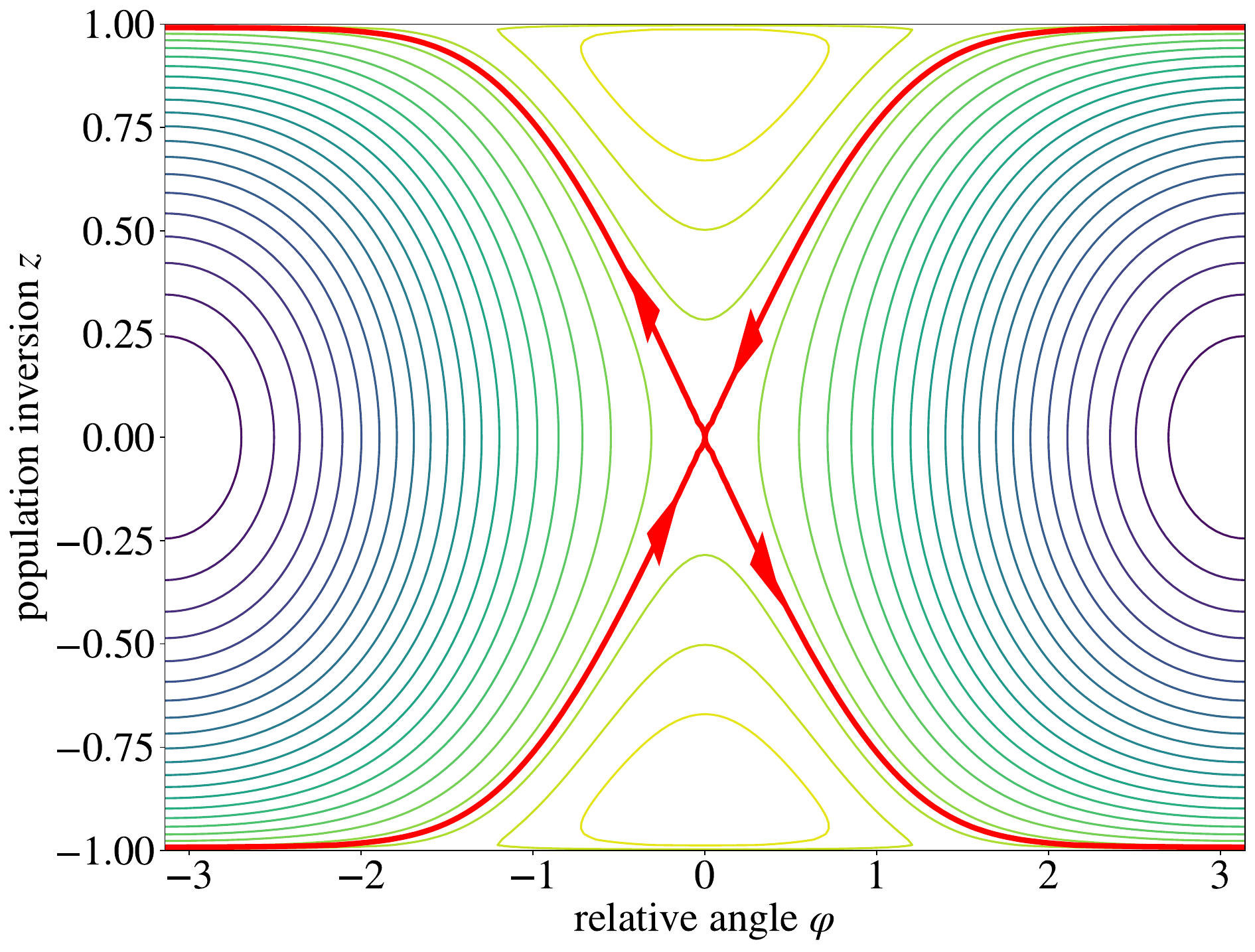}
    \caption{Reduced phase space structure ($z,\varphi)$ for $\Theta_{\ast}$: contour lines of the Hamiltonian correspond to the classical trajectories. There are %three stable fixed 
    three elliptic fixed points and one hyperbolic fixed point whose associated separatrix is highlighted in red. Arrows on the separatrix indicate the stable and unstable manifolds.}
    \label{fig:PhaseSpaceFP}
\end{figure}

Fig.~\ref{fig:PhaseSpaceFP} shows the reduced phase space structure of the system, generated by Eq.~\eqref{eq:EquationOfMotion}, at the maximal unstable parameter $\Theta_{\ast}$. 
Note in particular the (red) separatrix defined by the unstable and stable manifold originating from the off-phase FP. 
The merging of stable and unstable manifolds indicates that the linearized regime is bounded, namely, any classical trajectory on the unstable manifolds converges to the stable manifold leading (in infinite time) back to the hyperbolic FP.
The exact size of the linearized regime, modeled by the constant $c$ in the previous Sec.~\ref{sec:OTOC_theory}, plays a negligible role for $\hbarE\to 0$, since it is additive and $\hbarE$-independent constant in the leaking  time $\tLeak = \ls^{-1} \ln(c/\Delta u) \sim \ls^{-1} \ln(1/\hbarE))/2 + \ls^{-1}\ln c $ in Eq.~\eqref{eq:WignerWeylFPLeaking} with $\Delta u \sim \hbarE^{1/2}$ (for a coherent state). Therefore we leave $c$ unspecified in the subsequent discussion.

%We verified that, as shown in  Fig.~\ref{fig:PhaseSpaceFP} for $\Theta_{\ast}$, we verify that this is the only unstable FP in the classical mean-field limit. This is also true for the whole range $\Theta\in (\arctan 2,\pi/2)$, where the antihom. FP is unstable.

Armed with this very specific phase-space structure, we carry an in-depth analytical study of the OTOC~$\mathbf{C}(t)$ for the dimer in the next section.

\subsection{Microscopic approach: separatrix dynamics} 
\label{sec:peter}
\ifthenelse{1=1}
{}
{
In this section we provide a microscopic understanding of the dynamical transition by means of the exact separatrix dynamics of the mean field limit.

In this section, we work out an analytical expression for the classical OTOC
associated with a quantum state that is launched on the separatrix point of the
Bose-Hubbard dimer.
The latter is described by the quantum Hamiltonian
\begin{equation}
  \hat{H} = - J \left( \hat{a}_1^\dagger \hat{a}_2 + \hat{a}_2^\dagger \hat{a}_1
  \right) + \frac{U}{2} \sum_{l=1,2} \hat{a}_l^\dagger\hat{a}_l^\dagger
  \hat{a}_l\hat{a}_l
\end{equation}
with $\hat{a}_l^\dagger,\hat{a}_l$ the bosonic particle creation and
annihilation operator on the site $l=1,2$,
$J$ the hopping parameter, and $U$ the interaction parameter.

This quantum system has a classical counterpart which is given in terms of
the discrete nonlinear Schr\"odinger equation (setting $\hbar = 1$)
\begin{eqnarray}
  i \frac{d \psi_1}{d t} & = & - J \psi_2 + U |\psi_1|^2 \psi_1 \,, \\
  i \frac{d \psi_2}{d t} & = & - J \psi_1 + U |\psi_2|^2 \psi_2 \,,
\end{eqnarray}
with $\psi_l$ the classical field amplitude that is associated with the site
$l=1,2$.
Expressing $\psi_l = \sqrt{n_l}e^{i \varphi_l}$ in terms of the (real)
classical site occupancies $n_l$ and phases $\varphi_l$, and performing
the canonical transformation
$(n1,n2,\varphi_1,\varphi_2) \mapsto (N,n,\phi,\varphi)$
to new classical variables defined through
\begin{eqnarray}
  N & = & n_1 + n_2 \,, \\
  n & = & \frac{1}{2}( n_1 - n_2) \,, \\
  \phi & = & \frac{1}{2}( \varphi_1 + \varphi_2 ) \,, \\
  \varphi & = & \varphi_1 - \varphi_2 - \pi \,,
\end{eqnarray}
we end up with a classical Hamiltonian system described the inter-site
population exchange dynamics via the Hamiltonian
\begin{equation}
  H(n,\varphi) = U n^2 + J \sqrt{N^2 - 4 n^2} \cos \varphi \,,
  \label{eq:BH.Hc}
\end{equation}
which parametrically depends on the constant of motion $N$ corresponding
to the total population of the system.
The classical time evolution of the system is given in terms of the
Hamiltonian equations of motion
\begin{eqnarray}
  \frac{d n}{d t} & = & \frac{\partial H}{\partial \varphi}(n,\varphi) =
  - J \sqrt{N^2 - 4 n^2} \sin \varphi \,, \label{eq:BH.n} \\
  \frac{d \varphi}{d t} & = & - \frac{\partial H}{\partial n}(n,\varphi) =
  - 2 U n + \frac{4 J n \cos \varphi}{\sqrt{N^2 - 4 n^2}} \,, \label{eq:BH.ph}
\end{eqnarray}
Expressed in terms of the relative population imbalance $z = 2 n / N$,
the rescaled dimensionless time $\tau = J t$, and the dimensionless
nonlinearity parameter
\begin{equation}
  \gamma = \frac{N U}{2 J} \,,
\end{equation}
Eqs.~\eqref{eq:BH.n} and \eqref{eq:BH.ph} are rewritten as
\begin{eqnarray}
  \dot{z} \equiv \frac{d z}{d \tau} & = & - 2 \sqrt{1 - z^2} \sin\varphi \,,
  \label{eq:BH.z} \\
  \dot{\varphi} \equiv \frac{d \varphi}{d \tau} & = & - 2 \gamma z
  + \frac{2 z \cos\varphi}{\sqrt{1 - z^2}} \label{eq:BH.p} \,.
\end{eqnarray}
}
We choose for the dimer the operators $\hat{A}=\hat{B}=\hat{n}_{1}$ to be the number operator $\hat{n}_{1}= \hat{a}^{\dagger}_{1} \hat{a}_{1}$ at the first site. Therefore, we get
\begin{align}
    \mathbf{C}(t) =\langle || [ \hat{n}_{1}(t), \hat{n}_{1} ] ||^2 \rangle = \langle || [ \hat{n}(t), \hat{n} ] ||^2 \rangle\, ,
    \label{eq:OTOC_defDimer}
\end{align} 
where $\hat{n} = \frac{1}{2}(\hat{n}_{1}-\hat{n}_{2})$. For this OTOC, we are interested in evaluating the classical expression which is
given by
\begin{equation}
  O(t) = \iint dn d\varphi\, W(n,\varphi)
  \left(\frac{\partial n_t}{\partial \varphi_0}\right)^2 \,, \label{eq:OTOCc}
\end{equation}
with $W(n,\varphi)$ the Wigner function associated with the initial state.
The latter is, for the sake of simplicity, modeled as a coherent quantum state
$\exp(\sqrt{N_0}(\hat{a}_-^\dagger - \hat{a}_-)) \ket{0}$ with $\hat{a}_{-}=\hat{a}_{1} -\hat{a}_{2}$, keeping in mind that for large
$N_0$ this coherent state features very similar properties as a
number-projected coherent state with total particle number $N_0$ as far as the
site population exchange dynamics is concerned.
The Wigner function associated with this initial state would be given by
\begin{align}
    \begin{split}
        W(N&,\phi,n,\varphi) \simeq 
        \\
        \frac{1}{\pi^2} &\exp\left(-
        \frac{(N - N_0)^2}{2 N_0} - 2 N_0 \phi^2 - \frac{2 n^2}{N_0} -
        \frac{N_0 \varphi^2}{2} \right)
    \end{split}
\end{align}
  in the framework of a quadratic expansion valid for $N_0 \gg 1$.
%(using again $\hbar = 1$).
Since the Wigner function $W$ describes a tight localization of $N$ about $N_0$, we set
$N_0 = N$ henceforth and model the initial quantum state concerning the
inter-site population exchange dynamics by the Wigner function
\begin{equation}
  W(n,\varphi) = \frac{1}{\pi} \exp\left(- \frac{2 n^2}{\omega N}
  -  \frac{N \omega \varphi^2}{2} \right) \,,
\end{equation}
where the squeezing parameter $\omega$ allows for some flexibility
in the definition of the initial quantum state.

Let us first discuss the linearized dynamics in the near vicinity of
the FP~$(n,\varphi) = (0,0)$.
Linearizing Eq.~\eqref{eq:EquationOfMotion}, we obtain
the system of equations 
\begin{align}
\begin{split}
     \dot{z} & =  - 4\cos \Theta \varphi \,,  \\
  \dot{\varphi} & = - 2 (\sin \Theta -2\cos{\Theta}) z \,, 
\end{split}
\label{eq:BH.peterlinearized}
\end{align}
which is readily solved as
\begin{align}
    \begin{split}
      z_t & = z_0 \cosh \ls t - \frac{4\cos{\Theta} \varphi_0}{\ls}
      \sinh \ls t\,,  \\
      \varphi_t & = \varphi_0\cosh \ls t - \frac{\ls z_0}{4\cos{\Theta}}
      \sinh \ls t 
    \end{split}
\label{eq:BH.peterlinearizTime}
\end{align}
in terms of the stability exponent
\begin{equation}
  \ls = 4\cos{\Theta}\sqrt{\frac{\gamma}{2} - 1} \, , \label{eq:BH.la}
\end{equation}
where we defined the %so called 
nonlinearity parameter $\gamma=\tan \Theta = gN/(LJ)$.
The latter becomes purely imaginary for $\gamma < 2$, which implies that
$(n,\varphi) = (0,0)$ turns into a stable fixed point if the nonlinearity
parameter $\gamma$ is decreased below two, as it is plotted in Fig.~\ref{fig:StablHomFP}.

Considering $\gamma > 2$ henceforth, and assuming that the point
$(z_0,\varphi_0)$ is located very closely to the origin in this phase space,
we can, as in the previous section, identify a time scale
$\tLoc \gg \ls^{-1}$ for which we still have $|z_{\tLoc}| \ll 1$ and
$|\varphi_{\tLoc}|\ll 1$, such that the above linearization Eq.~\eqref{eq:BH.peterlinearized} of the classical
equations of motion remains valid until $t = \tLoc$.
Since at the same time we have $\ls \tLoc \gg 1$ by assumption,
the solution Eq.~\eqref{eq:BH.peterlinearizTime} of the linearized
equation Eq.~\eqref{eq:BH.peterlinearized} for $t= \tLoc$
simplifies as
\begin{eqnarray}
  z_{\tLoc} & = & \left(\frac{z_0}{2} -  \frac{2\cos \Theta\varphi_0}{\ls}\right)
  e^{\ls \tLoc} \,, \label{eq:BH.za} \\
  \varphi_{\tLoc} & = & \left(\frac{\varphi_0}{2} - \frac{\ls z_0}{8 \cos \Theta}
  \right) e^{\ls \tLoc} \,.
\end{eqnarray}

From the time $\tLoc$ on, we can safely assume that the trajectory
under consideration very closely follows the separatrix structure emanating
from the unstable \resubmit{\emph{off-phase}} fixed point $(z,\varphi) =(0,0)$. 
This separatrix structure is obtained through the identification of the
energy
\begin{equation}
  H(n,\varphi) = 2\cos \Theta N + \sin{\Theta} \frac{N}{2}
\end{equation}
of the classical Hamiltonian Eq.~\eqref{eq:classicalHamiltonian}, from which follows the identity
\begin{equation*}
  \cos\varphi = \frac{1 - \frac{\gamma}{4} z^2 }{\sqrt{1 - z^2}} \,.
\end{equation*}
Inserting this expression into Eq.~\eqref{eq:EquationOfMotion} yields the differential
equation
\begin{equation*}
  \dot{z} = z \sqrt{\ls^2 - \sin^2\Theta  z^2}
\end{equation*}
describing the motion along the upper or lower separatrix branch.
This equation is straightforwardly integrated yielding
\begin{align*}
  t - \tLoc =& \int_{z_{\tLoc}}^{z_t}
  \frac{\d z}{z \sqrt{\ls^2 - \sin^2\Theta z^2}}
  \\
   =& - \frac{1}{\ls}\left[ \mathrm{arcosh}
    \left(\frac{\ls}{\sin \Theta |z_{t}|} \right) - \mathrm{arcosh}
    \left(\frac{\ls}{\sin \Theta |z_{\tLoc}|} \right) \right] \,,
\end{align*}
from which we obtain
\begin{equation}
  z_t= \frac{\mathrm{sgn}(z_{\tLoc}) \ls / \sin\Theta}
  {\cosh\left[\mathrm{arcosh} \left(\frac{\ls}{\sin\Theta |z_{\tLoc}|}
      \right) - \ls (t - \tLoc) \right]} \,.
      \label{eq:zt}
\end{equation}
Using $|z_{\tLoc}| \ll 1$ and hence also $\sin\Theta |z_{\tLoc}|/\ls \ll 1$
for finite values of $\sin\Theta$ and $\ls$, we define
\begin{align}
  \begin{split}
    x_t & = \mathrm{sgn}(z_{\tLoc})\exp\left[-\mathrm{arcosh}
    \left(\frac{\ls}{\gamma |z_{\tLoc}|}\right) + \ls (t - \tLoc)
    \right]  
    \\
  & \simeq  \frac{\sin \Theta}{2 \ls} \left(\frac{z_0}{2} -
  2\cos \Theta\frac{\varphi_0}{\ls}\right)e^{\ls t} 
  \\
  &\simeq
  \frac{\sin \Theta}{\ls} \left(\frac{n_0}{N} - 2\cos \Theta
  \frac{\varphi_0}{\ls}\right)\sinh(\ls t) \,, 
  \end{split}
  \label{eq:xt}
\end{align}
where we make use of the asymptotic expression
\begin{equation}
  \mathrm{arcosh}(u) = \ln\left(u + \sqrt{u^2 - 1}\right) \simeq \ln(2u)
   + O(u^{-2})
   \label{eq:arcosh}
\end{equation}
for large $u$, in combination with Eq.~\eqref{eq:BH.za}.
With $(\cosh u)^{-1} = 2 e^u / ( 1 + e^{2u})$ %together with Eqs.~\eqref{eq:arcosh} and \eqref{eq:xt}, 
we can thus rewrite Eq.~\eqref{eq:zt} in terms of the expression Eq.~\eqref{eq:xt} as
\begin{equation}
  z_\tau = \frac{2 \ls}{\sin \Theta} \frac{x_t}{1 + x_t^2} \,,
\end{equation}
which yields
\begin{equation}
  n_t = \frac{N \ls}{\sin \Theta} \frac{x_{t}}{1 + x_{t}^2}
  \label{eq:BH.nt} \,.
\end{equation}
Replacing $e^{\ls t}$ with $2 \sinh(\ls t)$ in Eq.~\eqref{eq:xt}
is clearly valid for large $\ls t\gg 1$ and has the additional advantage
that the short-time regime in the time evolution of $n_t$ will thereby be
correctly captured as well within Eq.~\eqref{eq:BH.nt}.
\resubmit{
With Eq.~\eqref{eq:BH.nt}, we write down the needed derivative 
\begin{align*}
    \frac{\partial n_{t}}{\partial \varphi_{0}} &= 
     \frac{N \ls}{\sin \Theta} \frac{1-x_{t}^{2}}{(1 + x_{t}^2)^{2}} \frac{\partial x_{t}}
     {\partial \varphi_{0}}
     \\
     &\simeq
     \frac{-2 \cos\Theta}{\ls}\sinh(\ls t)  \frac{1-x_{t}^{2}}{(1 + x_{t}^2)^{2}} .
\end{align*}}%
The classical limit of the quantum OTOC, Eq.~\eqref{eq:OTOCc},
is then evaluated by \resubmit{substituting $n_{0}$ by $x_{t}$ via Eq.~\eqref{eq:xt} and integrating out the Gaussian integral in $\varphi_{0}$, yielding } 
\begin{align}
    \begin{split}
    O(t) = \frac{2\cos^2\Theta N^2}{ \sqrt{\pi} a \ls^2}& \sinh(\ls t)
    \\
  \int\limits_{\resubmit{-\infty}}^{\resubmit{\infty}} \frac{(1 - x^2)^2}{(1 + x^2)^4}&
  \exp\left[-\left(\frac{ x}{2a \sinh(\ls t)}\right)^2\right] \d x 
    \end{split}
    \label{eq:exact.clOTOC}
\end{align}
where the dimensionless scale is defined as
\begin{equation}
  a = \frac{\sin{\Theta}/\ls}{\sqrt{8 \omega N}}
  \sqrt{\omega^2 + \frac{16 \cos^2 \Theta }{\ls^2}} \,.
  \label{eq:aParameter}
\end{equation}
% Using Eq.~\eqref{eq:BH.la}, this expression simplifies in the special
% case $\omega = 1$ yielding
% \begin{equation}
%   a = \frac{1}{\gamma - 1}\sqrt{\frac{\gamma^3}{32 N}}
%   = \frac{N \sqrt{U^3/J}}{8(N U - 2 J)} \,.
% \end{equation}
The short-time behavior of the OTOC, for $t \ll \tLeak = - \ls^{-1} \ln a$,
is yielded as
\begin{equation}
  O(t) \simeq 4\cos^{2} \Theta \frac{N^2}{\ls^2} \sinh^2(\ls t) \,,
  \label{eq:shortClassical}
\end{equation}
while for $t \gg \tLeak$ we obtain
\begin{equation}
  O(t) \simeq \cos^{2} \Theta \frac{\sqrt{\pi} N^2}{4 a \ls^2} e^{\ls t} \,.
  \label{eq:longClassical}
\end{equation}
These two limits correspond to the %heuristic derived 
$2\ls-\ls$ transition 
that we heuristically derived in Sec.~\ref{sec:OTOC_theory}, as expressed in 
Eq.~\eqref{eq:OverviewFP}. % in the previous Sec.~\ref{sec:OTOC_theory}. 
Note that the here defined $\tLeak$ agrees with the case ii) in Sec.~\ref{sec:OTOC_theory}.
The dimensionless constant $\ln a$ encodes the linear width of the wave-packet along the unstable direction. 

With this %extensive 
detailed classical calculation at hand, we analyze the OTOC centered around this local hyperbolic off-phase FP, considering the parameter $\Theta_{\ast}$ at which the value of the stability exponent is maximal, $\ls = 0.97$. 

\subsection{Numerical results for the Out-of-Time-Order Correlator}

We proceed now with the numerical study and calculate the OTOC via Eq.~\eqref{eq:OTOC_defDimer} by means of %extensive 
numerically exact simulations for the operators $\hat{A}=\hat{B}=\hat{n}_{1}$. We %will 
consider the state 
\begin{equation}
    |{\vec{\xi}}\,\rangle = \frac{1}{\mathcal{N}} \big( \xi_1 \hat{a}_1^{\dagger} + \xi_2 \hat{a}_2^{\dagger} \big)^{N} \ket{0}, \label{eq:cohst}
\end{equation}
which is a number-projected coherent state centered at the off-phase FP $\vec{\xi} = (\xi_1,\xi_2) = (\sqrt{N/2},-\sqrt{N/2})$, with $\mathcal{N} = \sqrt{N^N N!}$ a normalization constant.

For large total particle number $N$, the projected coherent state inherits properties from the coherent state, in particular the linear width of $\hbar_{{\rm eff}}^{1/2}$ in each phase space direction \cite{gardiner2004quantum}, including the unstable direction in Fig.~\ref{fig:PhaseSpaceFP}. Furthermore it sets the squeezing parameter $\omega=1$ in the classical analysis in Eq.~\eqref{eq:aParameter}. 
Following the discussion in Sec.~\ref{sec:OTOC_theory}, case ii), the leaking time $\tLeak$ is therefore half the Ehrenfest time $\tEhr$.
\begin{figure}[h!]
    \centering
    \includegraphics[width=\linewidth]{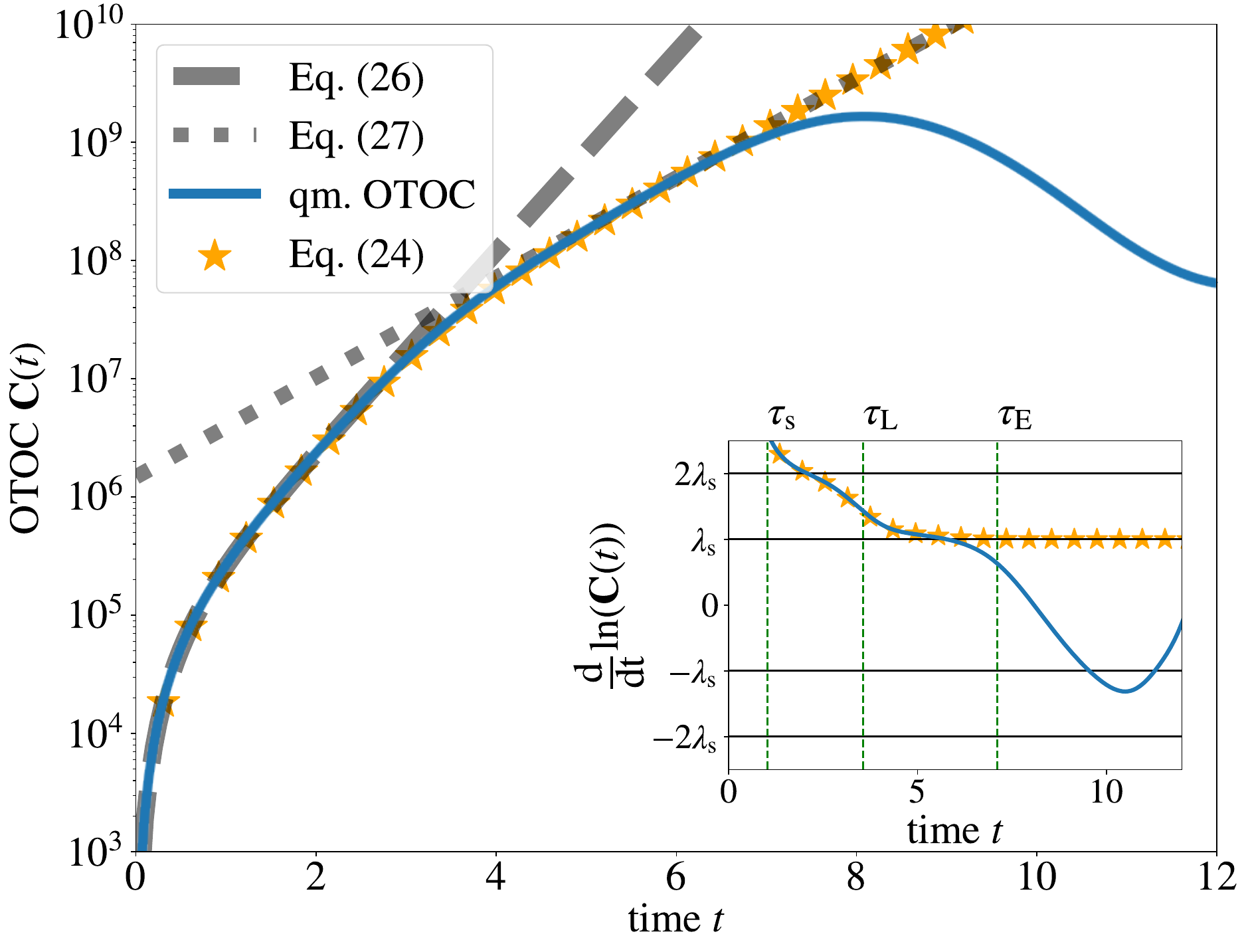}\\
    \includegraphics[width=\linewidth]{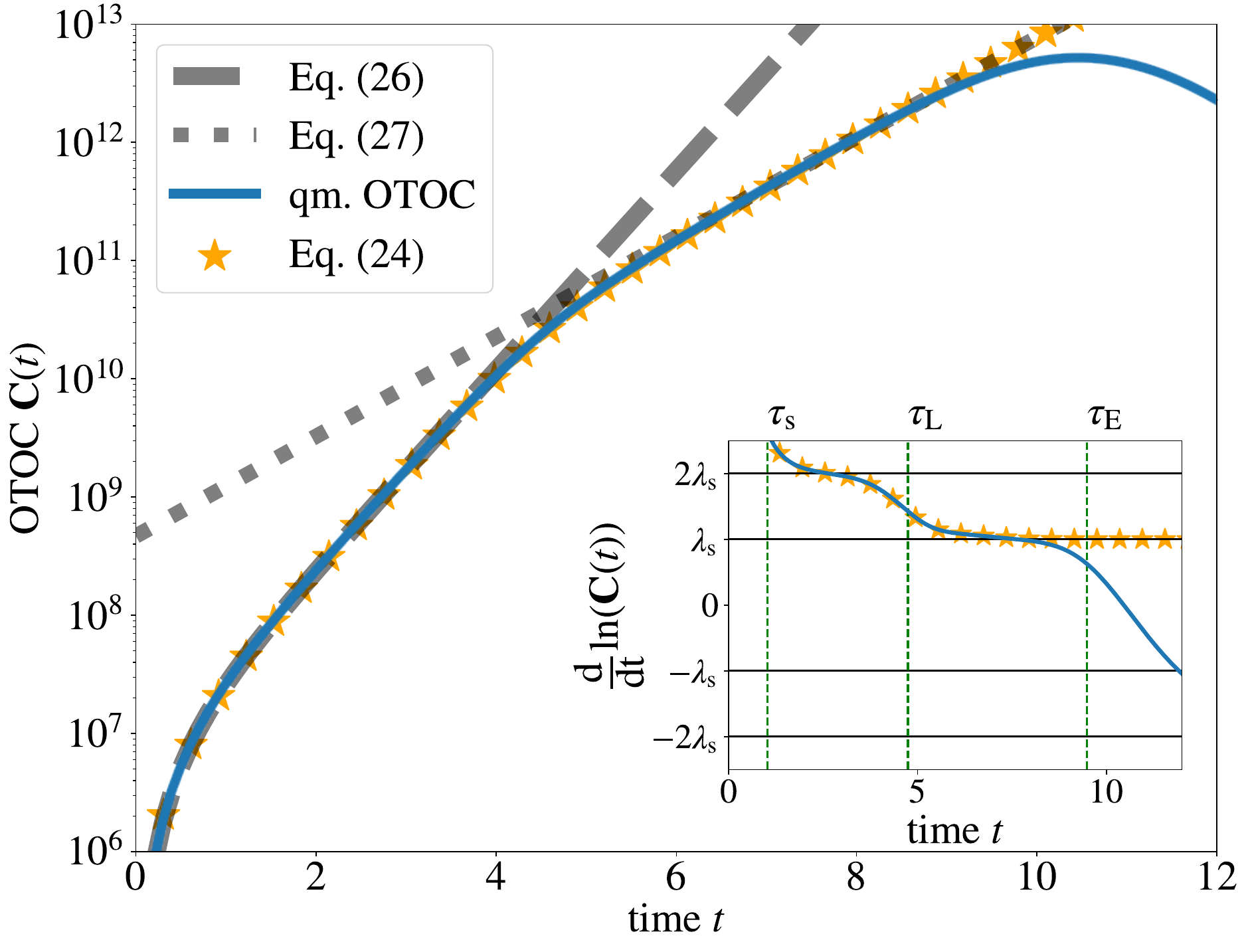}\\
    \includegraphics[width=\linewidth]{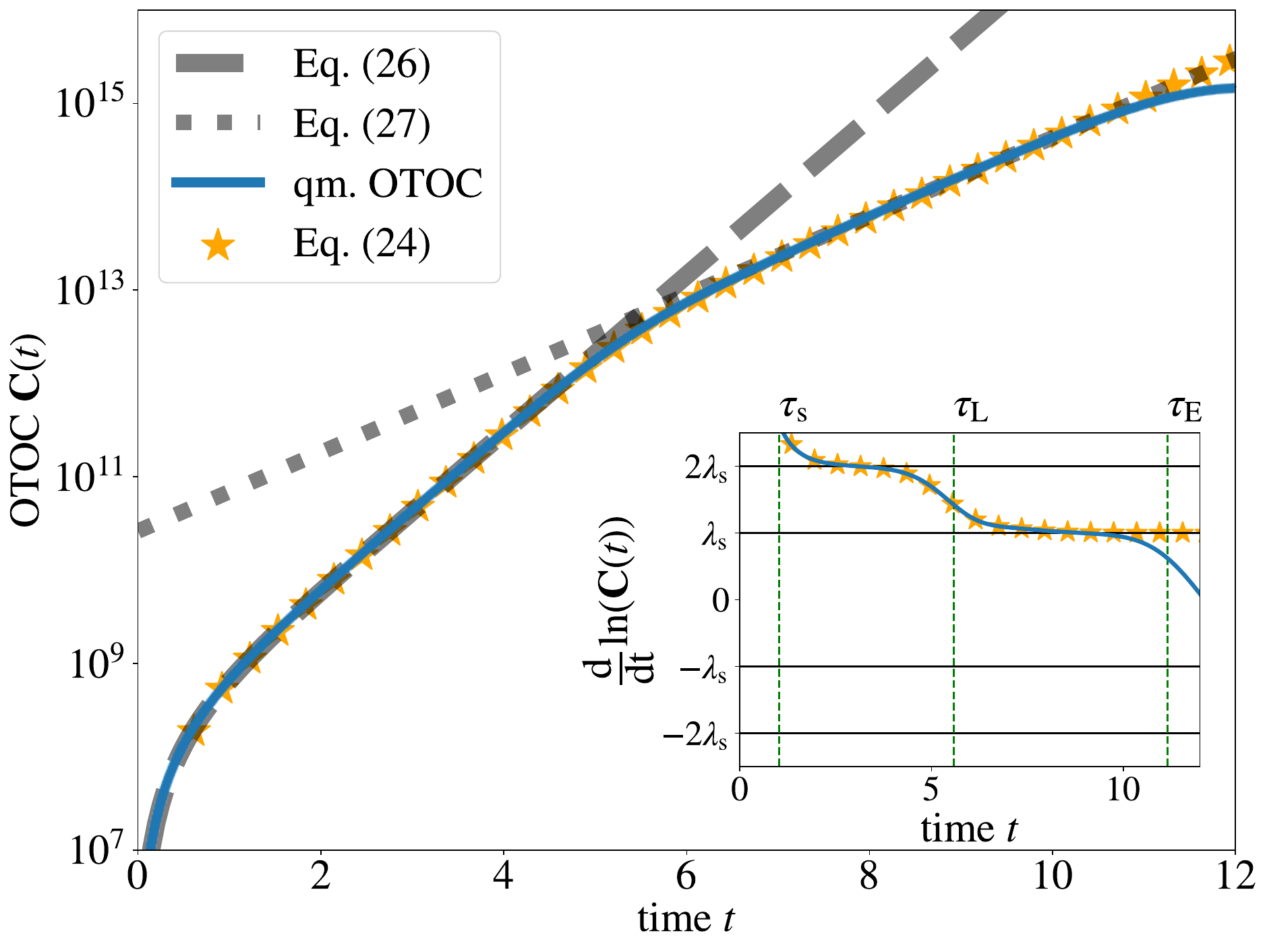}
    \caption{Top to bottom: OTOC $\mathbf{C}(t)$ for $N=10^3, ~10^4, ~5\cdot 10^4$ and $\Theta=1.35$; shaping kink from the $2\ls-\ls$ transition at $\tLeak=\tEhr/2$;  the classical expressions Eq.~\eqref{eq:shortClassical} and Eq.~\eqref{eq:longClassical} fit tightly the OTOC in each region showing $\exp(2\ls t)$ and $\exp(\ls t)$ exponential growth rates.}
    \label{fig:OTOC_FP_numeric}
\end{figure}
 We display our numerical OTOCs for increasing particle \resubmit{number} $N=10^3, ~10^4, ~5\cdot 10^4$ in Fig.~\ref{fig:OTOC_FP_numeric}, where we observe the predicted $2\ls-\ls$ transition, precisely following the heuristic arguments of Sec.~\ref{sec:OTOC_theory} and the %exact 
 more refined classical analysis of Sec.~\ref{sec:peter}. In particular, the analytical result Eq.~\eqref{eq:exact.clOTOC} for the classical OTOC follows \resubmit{very well} the quantum OTOC $\mathbf{C}(t)$, i.e., it captures both regimes and the transition.
 The kink at the transition becomes sharper for $N\to \infty$.  
 This is also seen in the insets showing the time-derivative of $\log (\mathbf{C}(t))$, which confirm that with increasing $N$ there are more and more pronounced $2\ls$ and $\ls$ regions of exponential growth.

\begin{figure*}[!ht]
    \centering
    \includegraphics[width=\linewidth]{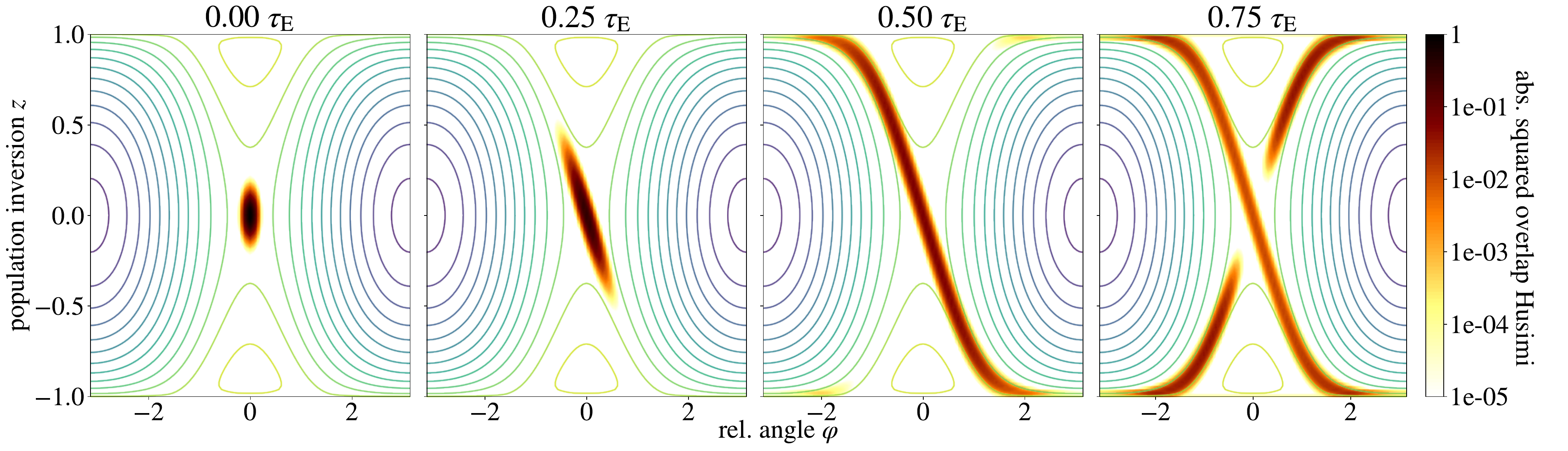}
    \caption{Time-evolution of the Husimi-distribution for the state $|\vec{\xi}\, \rangle$ centered at the hyperbolic off-phase FP with $N=10^3$ particles. We observe scrambling along the unstable manifold on the separatrix until $t\approx \tLeak$. \resubmit{The thin lines indicate the classical energy contours to visualize the spreading along the the unstable manifold of the \emph{off-phase} FP.}}
    \label{fig:husimiFP}
\end{figure*}
In order to illustrate 
the leaking from the linearized region around the FP, we visualize  in Fig.~\ref{fig:husimiFP}, via its Husimi distribution, the time
evolution of a wave packet that emanates from the coherent state Eq.~\eqref{eq:cohst} with $N=10^{3}$. 
With time, the wave packet spreads along the separatrix and evolves to the upper right and lower left corners of the phase space. 
We see that at the time $\tLeak = \tEhr/2$ the wave packet folds back from the unstable to the stable manifold. 
This back-folding corresponds to the dynamical \resubmit{crossover} of leaking from the linearized regime.

%The excellent agreement between our physical picture based on the leaking mechanism and the numerical simulations opens the possibility of manipulating the leaking time $\tLeak$, and therefore the different scrambling regimes, via squeezing the initial state $|{\vec{\xi}}\,\rangle$, as we discuss next.

\begin{figure*}[]
    \centering
    \includegraphics[width=0.5\linewidth]{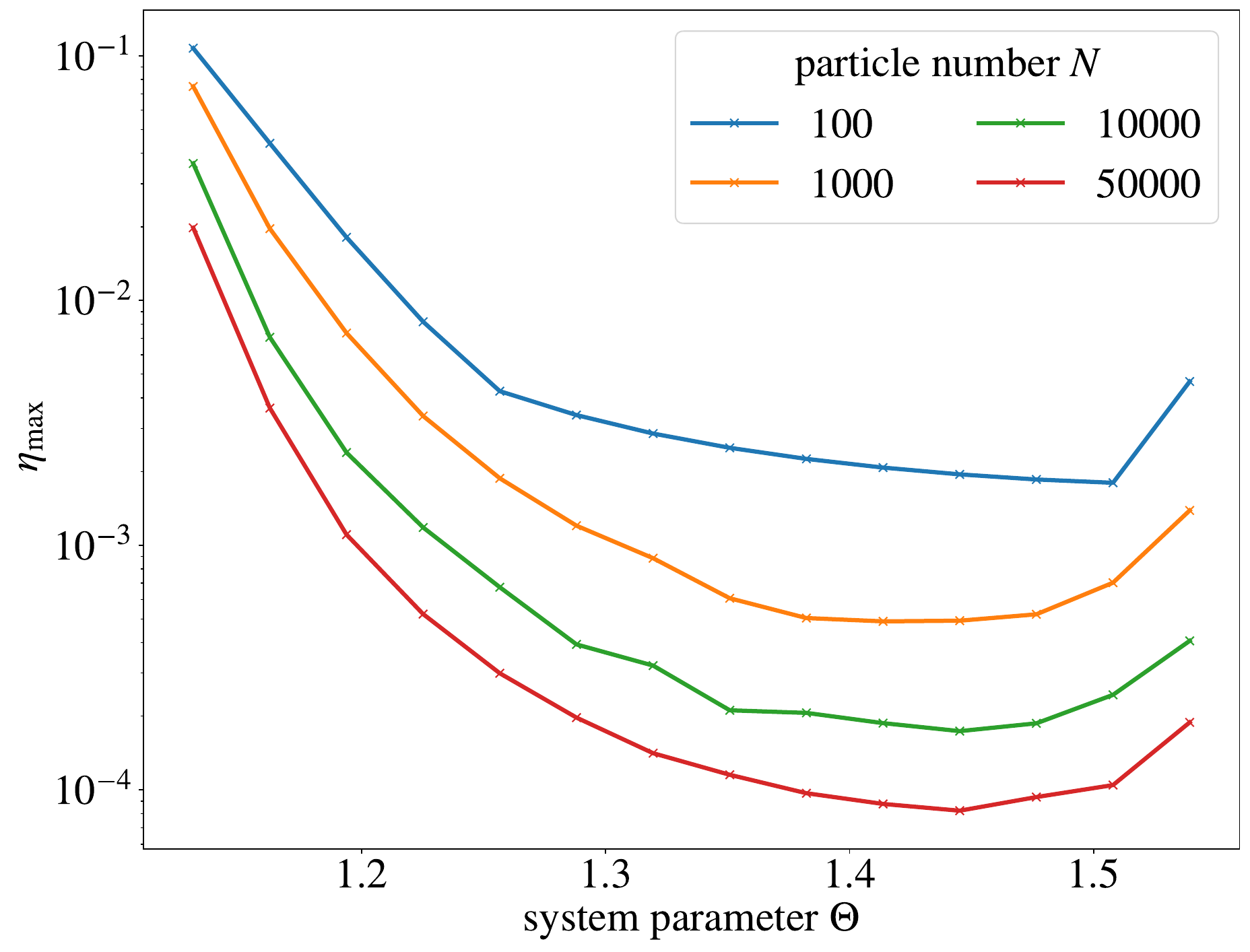}%
    \includegraphics[width=0.5\linewidth]{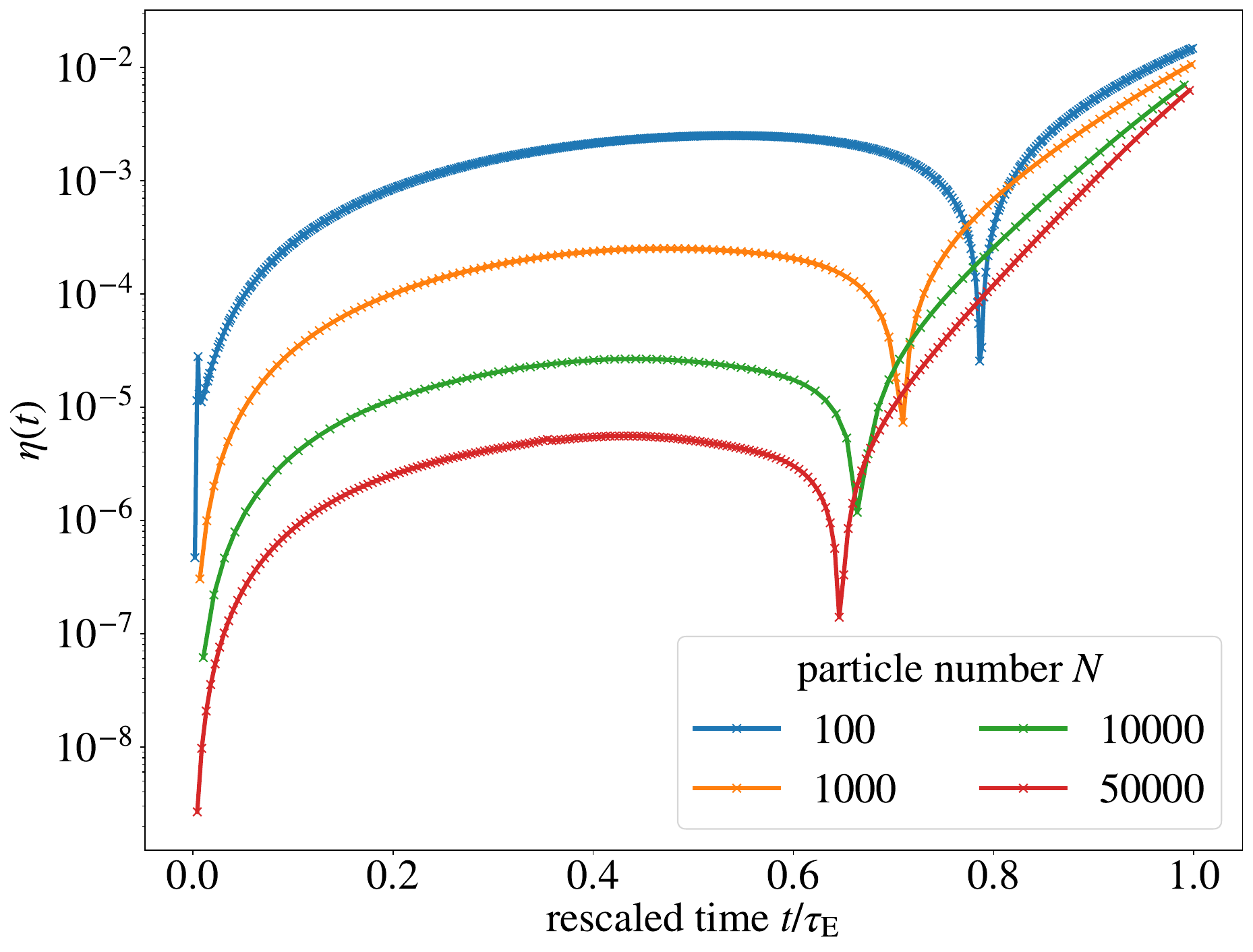}%
    \caption{Left panel: maximal relative logarithmic deviation Eq.~\eqref{eq:etamax} plotted as a function of the system parameter $\Theta$ in the instability region. Right panel: relative logarithmic deviation Eq.~\eqref{eq:deviationPrediction} plotted as a function of time for $\Theta=1.35$. The calculations were done for the total particle numbers (from top to bottom) $N = 100$, $1000$, $10000$, and $50000$.}
    %\caption{Fitted exponents for the $2\ls$ (upper panel) and the~$\ls$ (lower panel) region; for $N\to \infty$, we see an increasing agreement with the classical predictions from Eq.~\eqref{eq:OverviewFP}.}
    \label{fig:parameterScan}
\end{figure*}

Calculations for different values of the system parameter~$\Theta$ yield qualitatively similar behavior in the range where the off-phase FP is unstable, again with excellent agreement between the quantum OTOC Eq.~\eqref{eq:OTOC_defDimer} and its classical approximation Eq.~\eqref{eq:exact.clOTOC}.
To demonstrate this we
compute the 
relative logarithmic deviation between Eq.~\eqref{eq:exact.clOTOC} and Eq.~\eqref{eq:OTOC_defDimer}, %i.e., we show the maximum of 
defined by
\begin{equation}
    \eta(t) = \Big|\frac{\ln O(t) - \ln \mathbf{C}(t)}{\ln O(t)}\Big|
    \label{eq:deviationPrediction}
\end{equation}
%in Fig. \ref{fig:parameterScan}
for the number projected state $|{\vec{\xi}}\,\rangle$. % in the time-window~$t<0.8 \tEhr$. 
Fig. \ref{fig:parameterScan} displays, for various choices of the total particle number $N$, the time evolution of $\eta$ up to the Ehrenfest time for $\Theta=1.35$ (right panel) as well as its maximal value
\begin{equation}
\eta_{\rm max} = \max\limits_{t \leq 0.8 \tEhr} \eta(t) \label{eq:etamax}
\end{equation}
in the interval $0  \leq t \leq 0.8\tEhr$ as a function of $\Theta$ (left panel).
%Together we show this quantity over time for the $\Theta=1.35$ in order to argue that the cutoff $t<0.8 \tEhr$ is justified, since at the local Ehrenfest time we see an expected increasing deviation. 
At the edges of the instability region (i.e., for $\Theta \to \arctan(2)$ or $\pi/2$) %of the instability ($\ls\to 0$) 
the maximal deviation significantly increases, since there we have $\ls\to0$ and thus cannot, for finite $\hbarE$, justify the assumption that the time evolution of all phase-space points covered by the initial Wigner function very closely follows the separatrix arc. 
Nevertheless, the deviations clearly tend to zero in the semiclassical limit $\hbarE \to 0$, independently of the values of $\Theta \in (\arctan(2),\pi/2)$ and $t\in[0,\tEhr]$. %the classical expression~Eq.~\eqref{eq:exact.clOTOC} describes the time evolution up to the Ehrenfest time $\tEhr$.
This confirms that the~$2\ls-\ls$ transition is independent of the specific value of~$\Theta$, i.e., it is a robust signal of a dynamical \resubmit{crossover}.
%\FloatBarrier

\subsection{Squeezing -- engineering the leaking time $\tLeak$}

%\rc{Jesus Christ, what the hell means "A further manifest of our theory"?}
An important consequence of the leaking mechanism is that the linear %width 
extent
of the initial state along the unstable manifolds is the key ingredient for the exact position of the $2\ls-\ls$ transition. 
In order to check this, %dependence, 
we %proceed to 
squeeze the coherent state on the off-phase FP and subsequently calculate the OTOC. 
%Interestingly, 

\begin{figure*}[th!]
    \centering
    \includegraphics[width=0.49\linewidth,page=6]{timeShift_antihom_Theta_1_35_N_1000_L_2.pdf}%
    \includegraphics[width=0.49\linewidth]{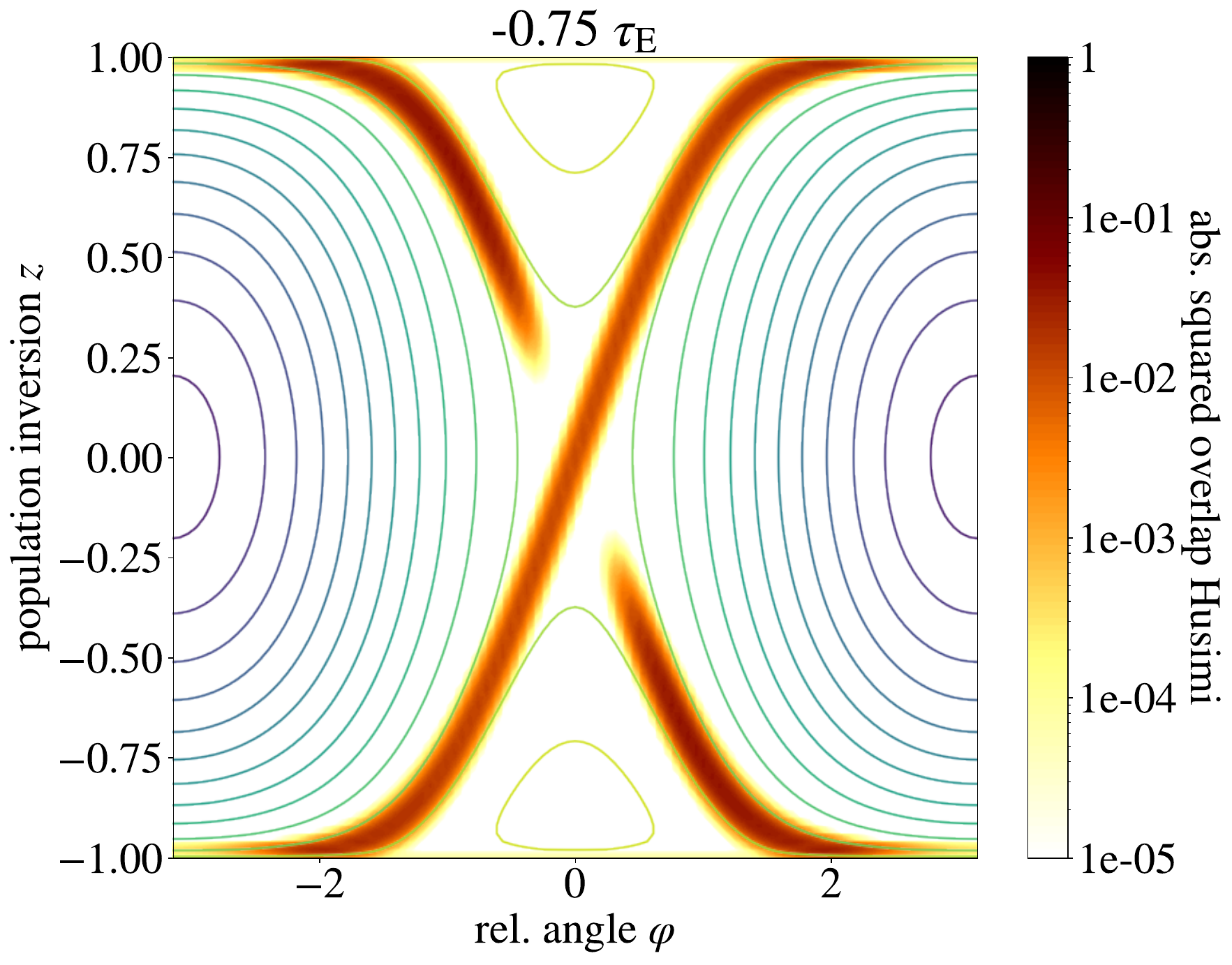}
    \caption{Left panel: OTOC for a squeezed coherent state for the system parameter $\Theta=1.35$ and particle number $N=10^{3}$. The \resubmit{crossover} to $\ls$ vanishes. Right panel: Husimi distribution of the squeezed coherent state. The state is distributed along the stable manifold of the off-phase FP and features tight localization, with a width $\sim \hbarE$, along the unstable manifold. }
    \label{fig:OTOC_FP_numeric_squeezed}
\end{figure*}

As a matter of fact,
squeezed states in optical lattices can be achieved experimentally to an exquisite degree \cite{squeezingBEC2008}. 
%In our theoretical setup, 
A squeezing protocol that is convenient for our purpose can be effectively (and unitarily) realized by reversing the time evolution, which in the experimental practice would amount to forward time propagation with reversed signs of the hopping parameter and the interaction parameter (to be done by Floquet engineering \cite{Lignier2007,Kierig2008} combined with Feshbach tuning).
This means, we replace $\ket{\xi }$ by 
\begin{align*}
    \ket{\xi (t_{0})} = \hat{U}(t_{0})\ket{\xi}
\end{align*}
with $t_{0} = -3\tEhr/4$, where $\hat{U}(t_{0})$ is the time-evolution operator, and then  calculate the OTOC, Eq.~\eqref{eq:OTOC_def}, for the initial state $\hat{\rho}(t_{0}) = \ket{\xi (t_{0}) }\bra{\xi (t_{0})} $. The corresponding scrambling dynamics is shown in then the left panel of Fig. \ref{fig:OTOC_FP_numeric_squeezed} for $N=10^3$, while the right panel depicts the initial Husimi distribution of the squeezed state.
This backward-time evaluated coherent state has a reduced linear %width 
extent
along the unstable manifold.
\resubmit{If we choose} $t_{0} = -3\tEhr/4$ (with $\tEhr$ the Ehrenfest time for the non-squeezed coherent state), we effectively transform $\Delta u \sim \hbar_{{\rm eff}}^{1/2}$ into $\Delta u \sim \hbarE$.
Thus, the new leaking time $\tLeak^{\ast}$ is now right at the Ehrenfest time, and no $2\ls-\ls$~transition is expected to exist, as fully confirmed by the numerical simulations.

%%%%%%%%%%%%%%%%% end BH%%%%%%%%%%%%%%%%%%%%%%%%%

\section{Conclusion }%\& Outlook}

\label{sec:conclusion}

The OTOC associated with a wave packet that is localized around a hyperbolic fixed point in a quantum system with integrable classical (mean-field) limit undergoes a transition between different dynamical regimes, which is driven by a leaking mechanism of phase space volume along classical separatrices. If located within the pre-Ehrenfest time scale, this dynamical \resubmit{crossover} imprints a characteristic kink structure to the scrambling as measured by the exponential form of out-of-time-order correlators. Specifically, the exponential growth changes from $2\ls $ to $\ls$, and the kink develops for $\hbarE\to 0$, where $\ls$ is the stability exponent of the fixed point. We derived an analytical theory and showed how  this behavior is directly related to the classical limit of the out-of-time-order correlators when their time dependence is governed by the separatrix dynamics emerging around an unstable fixed point.  

Following this picture, we showed that squeezing the initial coherent state allows us to engineer the leaking time and thus the dynamical transition itself exactly as predicted by our analytical considerations. 

If the phase-space localization scale of the initial state is strong enough, the leaking time is beyond the Ehrenfest time and we obtain the standard $2\ls$ exponent. In contrast, an uniform state starts to leak immediately and even before the ergodic time. Therefore, the infinite temperature OTOC grows only with the reduced exponent~$\ls$.

We supported our picture of the dynamical \resubmit{crossover} by means of extensive simulations on the experimentally accessible, and integrable,  Bose-Hubbard dimer. The extremely clean fixed point and separatrix structure of this systems allows us for a detailed study of the mechanism, and the   analytical expectations of Sec.~\ref{sec:OTOC_theory} are verified to an excellent degree.  

In order to focus on separatrix effects like the leaking mechanism, requiring a very well controlled classical phase space, we restricted our numerical findings and the corresponding analytical theory to integrable systems for which the theory in Sec.~\ref{sec:OTOC_theory} assumes a bounded linearized regime around the fixed point. 
An ansatz generalizing these concepts to the realm of chaotic systems is that
the role of the stability exponent~$\ls$ from the fixed point is translated to the Lyapunov exponent~$\lambda_{\rm L}$ from the chaotic sea, thus providing a bridge to 
exponential growth laws of OTOCs that were found for quantum maps with the exponents $2\lambda_{\rm L}$ \cite{Rozenbaum_2017,argentinians} and $\lambda_{\rm L}$ \cite{arul2019}, respectively. A first numerical exploration of this matter, focusing on the behaviour of OTOCs in chaotic separatrix layers, was carried out in \cite{meier2023signatures}, where %the authors treat  the combination of unstable fixed points and chaotic layers investigating 
the transition from the stability exponent $\ls$ associated with an unstable fixed point to the Lyapunov exponent $\lambda_{\rm L}$ characterizing the chaotic layer was investigated for pre-Ehrenfest time scales. 
In such a situation, it is an open and thrilling question to which extent the $2\ls$-$\ls$ transition could distinguish a chaotic from an integrable system 
and can be generalized to mixed regular-chaotic dynamics.

\section{Acknowledgments}
We are grateful for financial support from the Deutsche Forschungsgemeinschaft (German
Research Foundation) through Project Ri681/15-1 (Project No. 456449460) within the Reinhart-Koselleck Programme.
MS further acknowledges funding through the Studienstiftung des Deutschen Volkes.

\section*{Appendix}

\subsection{Wigner-Moyal expansion}
\label{sec:appendixWignerMoyal} 
In this appendix we outline shortly how to obtain Eq.~\eqref{eq:WignerWeyl} via a Wigner-Moyal expansion. We refer for an introduction to the phase-space formalism to \cite{gardiner2004quantum,Case2008,Curtright2012}. 
We start with the exact phase space expression for the expectation value of a general operator $\hat{O}$ given in this representation by
\begin{align*}
    \langle \hat{O} \rangle &= \Tr \hat{\rho} \hat{O} = \int \!\!\int \d^{L} {q} \d^{L} {p} ~W_{\rho}(\Vec{q},\Vec{p}) O_{\rm W}(\Vec{q},\Vec{p})
\end{align*}
where $W_{\rho}$ is the Wigner function of a state described by the density operator $\hat{\rho}$, and $O_{\rm W}$ is the Wigner-Weyl symbol of the operator $\hat{O}$. Note here that $O_{\rm W}$ is a function, not an operator, i.e., $O_{\rm W}(\Vec{q},\Vec{p})$ is a complex number. 

To proceed, we will also need the Wigner-Weyl symbol of a product of operators $\hat{A}$ and $\hat{B}$, given by the so-called star product of the corresponding  Wigner-Weyl symbols as
\begin{align*}
    \big[ \hat{A}\hat{B}\big]_{\rm W}(\Vec{q},\Vec{p}) =& A_{\rm W}(\Vec{q},\Vec{p}) \star  B_{\rm W}(\Vec{q},\Vec{p})
    \intertext{where the star product is an abbreviation for }
    = A_{\rm W}(\Vec{q},\Vec{p}) 
    \operatorname{exp} \Big\{\frac{i\hbarE}{2} &(\overset{\!\longleftarrow}{\nabla_{\Vec{p}}} \cdot \overset{\!\longrightarrow}{\nabla_{\Vec{q}}} - \overset{\!\longleftarrow}{\nabla_{\Vec{q}}} \cdot \overset{\!\longrightarrow}{\nabla_{\Vec{p}}}) \Big\}
    B_{\rm W}(\Vec{q},\Vec{p})
    \intertext{and can be expanded in $\hbarE$ as}=  A_{\rm W}(\Vec{q},\Vec{p}) B_{\rm W}(\Vec{q},\Vec{p}) +& \frac{i\hbarE}{2} \big\{A_{\rm W}(\Vec{q},\Vec{p}), B_{\rm W}(\Vec{q},\Vec{p}) \big\} \\ &+ O(\hbarE^2).
\end{align*}
With this expression at hand, starting with Eq.~\eqref{eq:OTOC_def} we easily obtain the expression
\begin{align*}
    \mathbf{C}(t) = \int &\!\!\int  \d^{L} {q} \d^{L} {p} ~W_{\rho}(\Vec{q},\Vec{p}) \\
    &\big[[\hat{A}(t),\hat{B}]\big]_{\rm W} (\Vec{q},\Vec{p})\star  \big[[\hat{A}(t),\hat{B}]\big]_{\rm W} (\Vec{q},\Vec{p}),
\end{align*}
which up to this point of the manipulations is exact. Since our goal is to obtain the leading order (in the Wigner-Moyal sense) for the OTOC, the next step is to expand the two star products inside the commutator $\big[[\hat{A}(t),\hat{B}]\big]_{\rm W} $ up to its first non-vanishing order  
\begin{align*}
    \big[[\hat{A}(t),\hat{B}]\big]_{\rm W}  =& i \hbarE \big\{ [\hat{A}(t)]_{\rm W}(\Vec{q},\Vec{p}),B_{\rm W}(\Vec{q},\Vec{p})  \big\} 
    \\ &+ O(\hbarE^{2}).
\end{align*}

As it can be derived from expanding the Heisenberg equation of motion 
\begin{align*}
    \frac{\d}{\d t} \hat{A} (t) = \frac{i}{\hbarE} \big[\hat{H},\hat{A}(t)\big ] \,,
\end{align*}
the time evolution of the Wigner-Weyl symbol of the operator $\hat{A}$ is given by
\begin{eqnarray*}
    [\hat{A}(t)]_{\rm W}(\Vec{q},\Vec{p}) &=& A_{\rm W}(\Vec{q}(\Vec{q},\Vec{p},t),\Vec{p}(\Vec{q},\Vec{p},t)) + O(\hbarE), \\ 
    &\eqqcolon& A_{\rm W}(\Vec{q},\Vec{p},t) + O(\hbarE),
\end{eqnarray*}
where $\Vec{q}(\Vec{q},\Vec{p},t)$ and $\Vec{p}(\Vec{q},\Vec{p},t)$ are the classical evolved phase space points.
One then can verify that the in leading  order the quantum and classical equations of motion agree since the quantum commutator reduces to the classical Poisson-bracket.
Putting all together, we arrive to the leading order result we in Eq.~\eqref{eq:WignerWeyl}
\begin{align*}
     \mathbf{ C}(t) \,=\,& \hbarE^{2} \langle W_{\rho}(\Vec{q}_{0},\Vec{p}_{0}) \big|\big\{ A_{\rm W}(\Vec{q}_{0},\Vec{p}_{0},t), B_{\rm W}(\Vec{q}_{0},\Vec{p}_{0}) \big\}  \big|^{2}\rangle_{\text{PS}}
            \\
            &+ O(\hbarE^{3}),
\end{align*}
where we suppressed the phase-space integral into $\langle.\rangle_{\rm PS}$ and renamed $(\Vec{q},\Vec{p})$ to $(\Vec{q}_{0},\Vec{p}_{0})$.

\bibliography{apssamp}

%aipnum4-2.bst 2019-01-14 (MD) hand-edited version of apsrev4-1.bst
\begin{thebibliography}{60}%
\makeatletter
\providecommand \@ifxundefined [1]{%
 \@ifx{#1\undefined}
}%
\providecommand \@ifnum [1]{%
 \ifnum #1\expandafter \@firstoftwo
 \else \expandafter \@secondoftwo
 \fi
}%
\providecommand \@ifx [1]{%
 \ifx #1\expandafter \@firstoftwo
 \else \expandafter \@secondoftwo
 \fi
}%
\providecommand \natexlab [1]{#1}%
\providecommand \enquote  [1]{``#1''}%
\providecommand \bibnamefont  [1]{#1}%
\providecommand \bibfnamefont [1]{#1}%
\providecommand \citenamefont [1]{#1}%
\providecommand \href@noop [0]{\@secondoftwo}%
\providecommand \href [0]{\begingroup \@sanitize@url \@href}%
\providecommand \@href[1]{\@@startlink{#1}\@@href}%
\providecommand \@@href[1]{\endgroup#1\@@endlink}%
\providecommand \@sanitize@url [0]{\catcode `\\12\catcode `\$12\catcode
  `\&12\catcode `\#12\catcode `\^12\catcode `\_12\catcode `\%12\relax}%
\providecommand \@@startlink[1]{}%
\providecommand \@@endlink[0]{}%
\providecommand \url  [0]{\begingroup\@sanitize@url \@url }%
\providecommand \@url [1]{\endgroup\@href {#1}{\urlprefix }}%
\providecommand \urlprefix  [0]{URL }%
\providecommand \Eprint [0]{\href }%
\providecommand \doibase [0]{https://doi.org/}%
\providecommand \selectlanguage [0]{\@gobble}%
\providecommand \bibinfo  [0]{\@secondoftwo}%
\providecommand \bibfield  [0]{\@secondoftwo}%
\providecommand \translation [1]{[#1]}%
\providecommand \BibitemOpen [0]{}%
\providecommand \bibitemStop [0]{}%
\providecommand \bibitemNoStop [0]{.\EOS\space}%
\providecommand \EOS [0]{\spacefactor3000\relax}%
\providecommand \BibitemShut  [1]{\csname bibitem#1\endcsname}%
\let\auto@bib@innerbib\@empty
%</preamble>
\bibitem [{\citenamefont {Xu}\ and\ \citenamefont
  {Swingle}(2022)}]{SwingleTutorial}%
  \BibitemOpen
  \bibfield  {author} {\bibinfo {author} {\bibfnamefont {S.}~\bibnamefont
  {Xu}}\ and\ \bibinfo {author} {\bibfnamefont {B.}~\bibnamefont {Swingle}},\
  }\href@noop {} {\  (\bibinfo {year} {2022})},\ \Eprint
  {https://arxiv.org/abs/2202.07060} {arXiv:2202.07060 [quant-ph]} \BibitemShut
  {NoStop}%
\bibitem [{\citenamefont {Richter}, \citenamefont {Urbina},\ and\ \citenamefont
  {Tomsovic}(2022)}]{Richter2022}%
  \BibitemOpen
  \bibfield  {author} {\bibinfo {author} {\bibfnamefont {K.}~\bibnamefont
  {Richter}}, \bibinfo {author} {\bibfnamefont {J.~D.}\ \bibnamefont
  {Urbina}},\ and\ \bibinfo {author} {\bibfnamefont {S.}~\bibnamefont
  {Tomsovic}},\ }\href {https://doi.org/10.1088/1751-8121/ac9e4e} {\bibfield
  {journal} {\bibinfo  {journal} {Journal of Physics A: Mathematical and
  Theoretical}\ }\textbf {\bibinfo {volume} {55}},\ \bibinfo {pages} {453001}
  (\bibinfo {year} {2022})}\BibitemShut {NoStop}%
\bibitem [{\citenamefont {Maldacena}, \citenamefont {Shenker},\ and\
  \citenamefont {Stanford}(2016)}]{Maldacena_2016}%
  \BibitemOpen
  \bibfield  {author} {\bibinfo {author} {\bibfnamefont {J.}~\bibnamefont
  {Maldacena}}, \bibinfo {author} {\bibfnamefont {S.~H.}\ \bibnamefont
  {Shenker}},\ and\ \bibinfo {author} {\bibfnamefont {D.}~\bibnamefont
  {Stanford}},\ }\href {https://doi.org/10.1007%2Fjhep08%282016%29106}
  {\bibfield  {journal} {\bibinfo  {journal} {Journal of High Energy Physics}\
  }\textbf {\bibinfo {volume} {2016}} (\bibinfo {year} {2016})}\BibitemShut
  {NoStop}%
\bibitem [{\citenamefont {Hayden}\ and\ \citenamefont
  {Preskill}(2007)}]{Hayden_2007}%
  \BibitemOpen
  \bibfield  {author} {\bibinfo {author} {\bibfnamefont {P.}~\bibnamefont
  {Hayden}}\ and\ \bibinfo {author} {\bibfnamefont {J.}~\bibnamefont
  {Preskill}},\ }\href {https://doi.org/10.1088/1126-6708/2007/09/120}
  {\bibfield  {journal} {\bibinfo  {journal} {Journal of High Energy Physics}\
  }\textbf {\bibinfo {volume} {2007}},\ \bibinfo {pages} {120} (\bibinfo {year}
  {2007})}\BibitemShut {NoStop}%
\bibitem [{\citenamefont {Sekino}\ and\ \citenamefont
  {Susskind}(2008)}]{Sekino:2008he}%
  \BibitemOpen
  \bibfield  {author} {\bibinfo {author} {\bibfnamefont {Y.}~\bibnamefont
  {Sekino}}\ and\ \bibinfo {author} {\bibfnamefont {L.}~\bibnamefont
  {Susskind}},\ }\href {https://doi.org/10.1088/1126-6708/2008/10/065}
  {\bibfield  {journal} {\bibinfo  {journal} {Journal of High Energy Physics}\
  }\textbf {\bibinfo {volume} {2008}},\ \bibinfo {pages} {065} (\bibinfo {year}
  {2008})}\BibitemShut {NoStop}%
\bibitem [{\citenamefont {Gu}\ and\ \citenamefont {Kitaev}(2019)}]{Kitaev2019}%
  \BibitemOpen
  \bibfield  {author} {\bibinfo {author} {\bibfnamefont {Y.}~\bibnamefont
  {Gu}}\ and\ \bibinfo {author} {\bibfnamefont {A.}~\bibnamefont {Kitaev}},\
  }\href {https://doi.org/10.1007/jhep02(2019)075} {\bibfield  {journal}
  {\bibinfo  {journal} {Journal of High Energy Physics}\ }\textbf {\bibinfo
  {volume} {2019}} (\bibinfo {year} {2019}),\
  10.1007/jhep02(2019)075}\BibitemShut {NoStop}%
\bibitem [{\citenamefont {Kobrin}\ \emph {et~al.}(2021)\citenamefont {Kobrin},
  \citenamefont {Yang}, \citenamefont {Kahanamoku-Meyer}, \citenamefont
  {Olund}, \citenamefont {Moore}, \citenamefont {Stanford},\ and\ \citenamefont
  {Yao}}]{standfordOTOC2021}%
  \BibitemOpen
  \bibfield  {author} {\bibinfo {author} {\bibfnamefont {B.}~\bibnamefont
  {Kobrin}}, \bibinfo {author} {\bibfnamefont {Z.}~\bibnamefont {Yang}},
  \bibinfo {author} {\bibfnamefont {G.~D.}\ \bibnamefont {Kahanamoku-Meyer}},
  \bibinfo {author} {\bibfnamefont {C.~T.}\ \bibnamefont {Olund}}, \bibinfo
  {author} {\bibfnamefont {J.~E.}\ \bibnamefont {Moore}}, \bibinfo {author}
  {\bibfnamefont {D.}~\bibnamefont {Stanford}},\ and\ \bibinfo {author}
  {\bibfnamefont {N.~Y.}\ \bibnamefont {Yao}},\ }\href
  {https://doi.org/10.1103/PhysRevLett.126.030602} {\bibfield  {journal}
  {\bibinfo  {journal} {Phys. Rev. Lett.}\ }\textbf {\bibinfo {volume} {126}},\
  \bibinfo {pages} {030602} (\bibinfo {year} {2021})}\BibitemShut {NoStop}%
\bibitem [{\citenamefont {Tsuji}\ and\ \citenamefont
  {Werner}(2019)}]{OTOCfermions}%
  \BibitemOpen
  \bibfield  {author} {\bibinfo {author} {\bibfnamefont {N.}~\bibnamefont
  {Tsuji}}\ and\ \bibinfo {author} {\bibfnamefont {P.}~\bibnamefont {Werner}},\
  }\href {https://doi.org/10.1103/PhysRevB.99.115132} {\bibfield  {journal}
  {\bibinfo  {journal} {Phys. Rev. B}\ }\textbf {\bibinfo {volume} {99}},\
  \bibinfo {pages} {115132} (\bibinfo {year} {2019})}\BibitemShut {NoStop}%
\bibitem [{\citenamefont {{Larkin}}\ and\ \citenamefont
  {{Ovchinnikov}}(1969)}]{classic}%
  \BibitemOpen
  \bibfield  {author} {\bibinfo {author} {\bibfnamefont {A.~I.}\ \bibnamefont
  {{Larkin}}}\ and\ \bibinfo {author} {\bibfnamefont {Y.~N.}\ \bibnamefont
  {{Ovchinnikov}}},\ }\href
  {http://jetp.ras.ru/cgi-bin/e/index/e/28/6/p1200?a=list} {\bibfield
  {journal} {\bibinfo  {journal} {Soviet Journal of Experimental and
  Theoretical Physics}\ }\textbf {\bibinfo {volume} {28}},\ \bibinfo {pages}
  {1200} (\bibinfo {year} {1969})}\BibitemShut {NoStop}%
\bibitem [{Note1()}]{Note1}%
  \BibitemOpen
  \bibinfo {note} {{Although the exponential behavior of OTOCs is a hallmark of
  classical dynamical stability and therefore present in single- and few-body
  systems as well, its use as a measure of the scrambling of correlations is
  appropriate only in the many-body context \cite
  {SwingleTutorial,Maldacena_2016}}}\BibitemShut {NoStop}%
\bibitem [{\citenamefont {Swingle}\ \emph {et~al.}(2016)\citenamefont
  {Swingle}, \citenamefont {Bentsen}, \citenamefont {Schleier-Smith},\ and\
  \citenamefont {Hayden}}]{proposalExperimentOTOC}%
  \BibitemOpen
  \bibfield  {author} {\bibinfo {author} {\bibfnamefont {B.}~\bibnamefont
  {Swingle}}, \bibinfo {author} {\bibfnamefont {G.}~\bibnamefont {Bentsen}},
  \bibinfo {author} {\bibfnamefont {M.}~\bibnamefont {Schleier-Smith}},\ and\
  \bibinfo {author} {\bibfnamefont {P.}~\bibnamefont {Hayden}},\ }\href
  {https://link.aps.org/doi/10.1103/PhysRevA.94.040302} {\bibfield  {journal}
  {\bibinfo  {journal} {Phys. Rev. A}\ }\textbf {\bibinfo {volume} {94}},\
  \bibinfo {pages} {040302} (\bibinfo {year} {2016})}\BibitemShut {NoStop}%
\bibitem [{\citenamefont {Gärttner}\ \emph {et~al.}(2017)\citenamefont
  {Gärttner}, \citenamefont {Bohnet}, \citenamefont {Safavi-Naini},
  \citenamefont {Wall}, \citenamefont {Bollinger},\ and\ \citenamefont
  {Rey}}]{Garttner2017}%
  \BibitemOpen
  \bibfield  {author} {\bibinfo {author} {\bibfnamefont {M.}~\bibnamefont
  {Gärttner}}, \bibinfo {author} {\bibfnamefont {J.~G.}\ \bibnamefont
  {Bohnet}}, \bibinfo {author} {\bibfnamefont {A.}~\bibnamefont
  {Safavi-Naini}}, \bibinfo {author} {\bibfnamefont {M.~L.}\ \bibnamefont
  {Wall}}, \bibinfo {author} {\bibfnamefont {J.~J.}\ \bibnamefont
  {Bollinger}},\ and\ \bibinfo {author} {\bibfnamefont {A.~M.}\ \bibnamefont
  {Rey}},\ }\href {https://doi.org/10.1038%2Fnphys4119} {\bibfield  {journal}
  {\bibinfo  {journal} {Nature Physics}\ }\textbf {\bibinfo {volume} {13}},\
  \bibinfo {pages} {781} (\bibinfo {year} {2017})}\BibitemShut {NoStop}%
\bibitem [{\citenamefont {Li}\ \emph {et~al.}(2017)\citenamefont {Li},
  \citenamefont {Fan}, \citenamefont {Wang}, \citenamefont {Ye}, \citenamefont
  {Zeng}, \citenamefont {Zhai}, \citenamefont {Peng},\ and\ \citenamefont
  {Du}}]{NMR_OTOC}%
  \BibitemOpen
  \bibfield  {author} {\bibinfo {author} {\bibfnamefont {J.}~\bibnamefont
  {Li}}, \bibinfo {author} {\bibfnamefont {R.}~\bibnamefont {Fan}}, \bibinfo
  {author} {\bibfnamefont {H.}~\bibnamefont {Wang}}, \bibinfo {author}
  {\bibfnamefont {B.}~\bibnamefont {Ye}}, \bibinfo {author} {\bibfnamefont
  {B.}~\bibnamefont {Zeng}}, \bibinfo {author} {\bibfnamefont {H.}~\bibnamefont
  {Zhai}}, \bibinfo {author} {\bibfnamefont {X.}~\bibnamefont {Peng}},\ and\
  \bibinfo {author} {\bibfnamefont {J.}~\bibnamefont {Du}},\ }\href
  {https://doi.org/10.1103/PhysRevX.7.031011} {\bibfield  {journal} {\bibinfo
  {journal} {Phys. Rev. X}\ }\textbf {\bibinfo {volume} {7}},\ \bibinfo {pages}
  {031011} (\bibinfo {year} {2017})}\BibitemShut {NoStop}%
\bibitem [{\citenamefont {Kidd}, \citenamefont {Safavi-Naini},\ and\
  \citenamefont {Corney}(2020)}]{dimerBH}%
  \BibitemOpen
  \bibfield  {author} {\bibinfo {author} {\bibfnamefont {R.~A.}\ \bibnamefont
  {Kidd}}, \bibinfo {author} {\bibfnamefont {A.}~\bibnamefont {Safavi-Naini}},\
  and\ \bibinfo {author} {\bibfnamefont {J.~F.}\ \bibnamefont {Corney}},\
  }\href {https://doi.org/10.1103/PhysRevA.102.023330} {\bibfield  {journal}
  {\bibinfo  {journal} {Phys. Rev. A}\ }\textbf {\bibinfo {volume} {102}},\
  \bibinfo {pages} {023330} (\bibinfo {year} {2020})}\BibitemShut {NoStop}%
\bibitem [{\citenamefont {Shen}\ \emph {et~al.}(2017)\citenamefont {Shen},
  \citenamefont {Zhang}, \citenamefont {Fan},\ and\ \citenamefont
  {Zhai}}]{Shen2017}%
  \BibitemOpen
  \bibfield  {author} {\bibinfo {author} {\bibfnamefont {H.}~\bibnamefont
  {Shen}}, \bibinfo {author} {\bibfnamefont {P.}~\bibnamefont {Zhang}},
  \bibinfo {author} {\bibfnamefont {R.}~\bibnamefont {Fan}},\ and\ \bibinfo
  {author} {\bibfnamefont {H.}~\bibnamefont {Zhai}},\ }\href
  {https://doi.org/10.1103/PhysRevB.96.054503} {\bibfield  {journal} {\bibinfo
  {journal} {Phys. Rev. B}\ }\textbf {\bibinfo {volume} {96}},\ \bibinfo
  {pages} {054503} (\bibinfo {year} {2017})}\BibitemShut {NoStop}%
\bibitem [{\citenamefont {Bohrdt}\ \emph {et~al.}(2017)\citenamefont {Bohrdt},
  \citenamefont {Mendl}, \citenamefont {Endres},\ and\ \citenamefont
  {Knap}}]{Bohrdt2017}%
  \BibitemOpen
  \bibfield  {author} {\bibinfo {author} {\bibfnamefont {A.}~\bibnamefont
  {Bohrdt}}, \bibinfo {author} {\bibfnamefont {C.~B.}\ \bibnamefont {Mendl}},
  \bibinfo {author} {\bibfnamefont {M.}~\bibnamefont {Endres}},\ and\ \bibinfo
  {author} {\bibfnamefont {M.}~\bibnamefont {Knap}},\ }\href
  {https://doi.org/10.1088/1367-2630/aa719b} {\bibfield  {journal} {\bibinfo
  {journal} {New Journal of Physics}\ }\textbf {\bibinfo {volume} {19}},\
  \bibinfo {pages} {063001} (\bibinfo {year} {2017})}\BibitemShut {NoStop}%
\bibitem [{\citenamefont {Rammensee}, \citenamefont {Urbina},\ and\
  \citenamefont {Richter}(2018)}]{Josef2018}%
  \BibitemOpen
  \bibfield  {author} {\bibinfo {author} {\bibfnamefont {J.}~\bibnamefont
  {Rammensee}}, \bibinfo {author} {\bibfnamefont {J.~D.}\ \bibnamefont
  {Urbina}},\ and\ \bibinfo {author} {\bibfnamefont {K.}~\bibnamefont
  {Richter}},\ }\href {https://doi.org/10.1103/PhysRevLett.121.124101}
  {\bibfield  {journal} {\bibinfo  {journal} {Phys. Rev. Lett.}\ }\textbf
  {\bibinfo {volume} {121}},\ \bibinfo {pages} {124101} (\bibinfo {year}
  {2018})}\BibitemShut {NoStop}%
\bibitem [{\citenamefont {Hummel}\ \emph {et~al.}(2019)\citenamefont {Hummel},
  \citenamefont {Geiger}, \citenamefont {Urbina},\ and\ \citenamefont
  {Richter}}]{Hummel2019}%
  \BibitemOpen
  \bibfield  {author} {\bibinfo {author} {\bibfnamefont {Q.}~\bibnamefont
  {Hummel}}, \bibinfo {author} {\bibfnamefont {B.}~\bibnamefont {Geiger}},
  \bibinfo {author} {\bibfnamefont {J.~D.}\ \bibnamefont {Urbina}},\ and\
  \bibinfo {author} {\bibfnamefont {K.}~\bibnamefont {Richter}},\ }\href
  {https://doi.org/10.1103/PhysRevLett.123.160401} {\bibfield  {journal}
  {\bibinfo  {journal} {Phys. Rev. Lett.}\ }\textbf {\bibinfo {volume} {123}},\
  \bibinfo {pages} {160401} (\bibinfo {year} {2019})}\BibitemShut {NoStop}%
\bibitem [{\citenamefont {Pappalardi}\ \emph {et~al.}(2018)\citenamefont
  {Pappalardi}, \citenamefont {Russomanno}, \citenamefont {\ifmmode
  \check{Z}\else \v{Z}\fi{}unkovi\ifmmode~\check{c}\else \v{c}\fi{}},
  \citenamefont {Iemini}, \citenamefont {Silva},\ and\ \citenamefont
  {Fazio}}]{Papparlardi2018}%
  \BibitemOpen
  \bibfield  {author} {\bibinfo {author} {\bibfnamefont {S.}~\bibnamefont
  {Pappalardi}}, \bibinfo {author} {\bibfnamefont {A.}~\bibnamefont
  {Russomanno}}, \bibinfo {author} {\bibfnamefont {B.}~\bibnamefont {\ifmmode
  \check{Z}\else \v{Z}\fi{}unkovi\ifmmode~\check{c}\else \v{c}\fi{}}}, \bibinfo
  {author} {\bibfnamefont {F.}~\bibnamefont {Iemini}}, \bibinfo {author}
  {\bibfnamefont {A.}~\bibnamefont {Silva}},\ and\ \bibinfo {author}
  {\bibfnamefont {R.}~\bibnamefont {Fazio}},\ }\href
  {https://doi.org/10.1103/PhysRevB.98.134303} {\bibfield  {journal} {\bibinfo
  {journal} {Phys. Rev. B}\ }\textbf {\bibinfo {volume} {98}},\ \bibinfo
  {pages} {134303} (\bibinfo {year} {2018})}\BibitemShut {NoStop}%
\bibitem [{\citenamefont {Xu}, \citenamefont {Scaffidi},\ and\ \citenamefont
  {Cao}(2020)}]{Scaffidi2020}%
  \BibitemOpen
  \bibfield  {author} {\bibinfo {author} {\bibfnamefont {T.}~\bibnamefont
  {Xu}}, \bibinfo {author} {\bibfnamefont {T.}~\bibnamefont {Scaffidi}},\ and\
  \bibinfo {author} {\bibfnamefont {X.}~\bibnamefont {Cao}},\ }\href
  {https://doi.org/10.1103/PhysRevLett.124.140602} {\bibfield  {journal}
  {\bibinfo  {journal} {Phys. Rev. Lett.}\ }\textbf {\bibinfo {volume} {124}},\
  \bibinfo {pages} {140602} (\bibinfo {year} {2020})}\BibitemShut {NoStop}%
\bibitem [{\citenamefont {Geiger}, \citenamefont {Urbina},\ and\ \citenamefont
  {Richter}(2021)}]{Geiger2021}%
  \BibitemOpen
  \bibfield  {author} {\bibinfo {author} {\bibfnamefont {B.}~\bibnamefont
  {Geiger}}, \bibinfo {author} {\bibfnamefont {J.~D.}\ \bibnamefont {Urbina}},\
  and\ \bibinfo {author} {\bibfnamefont {K.}~\bibnamefont {Richter}},\ }\href
  {https://doi.org/10.1103/PhysRevLett.126.110602} {\bibfield  {journal}
  {\bibinfo  {journal} {Phys. Rev. Lett.}\ }\textbf {\bibinfo {volume} {126}},\
  \bibinfo {pages} {110602} (\bibinfo {year} {2021})}\BibitemShut {NoStop}%
\bibitem [{\citenamefont {Villaseñor}\ \emph {et~al.}(2023)\citenamefont
  {Villaseñor}, \citenamefont {Pilatowsky-Cameo}, \citenamefont
  {Bastarrachea-Magnani}, \citenamefont {Lerma-Hernández}, \citenamefont
  {Santos},\ and\ \citenamefont {Hirsch}}]{Santos1}%
  \BibitemOpen
  \bibfield  {author} {\bibinfo {author} {\bibfnamefont {D.}~\bibnamefont
  {Villaseñor}}, \bibinfo {author} {\bibfnamefont {S.}~\bibnamefont
  {Pilatowsky-Cameo}}, \bibinfo {author} {\bibfnamefont {M.~A.}\ \bibnamefont
  {Bastarrachea-Magnani}}, \bibinfo {author} {\bibfnamefont {S.}~\bibnamefont
  {Lerma-Hernández}}, \bibinfo {author} {\bibfnamefont {L.~F.}\ \bibnamefont
  {Santos}},\ and\ \bibinfo {author} {\bibfnamefont {J.~G.}\ \bibnamefont
  {Hirsch}},\ }\href {https://doi.org/10.3390/e25010008} {\bibfield  {journal}
  {\bibinfo  {journal} {Entropy}\ }\textbf {\bibinfo {volume} {25}} (\bibinfo
  {year} {2023}),\ 10.3390/e25010008}\BibitemShut {NoStop}%
\bibitem [{\citenamefont {Benet}\ \emph {et~al.}(2022)\citenamefont {Benet},
  \citenamefont {Borgonovi}, \citenamefont {Izrailev},\ and\ \citenamefont
  {Santos}}]{Santos2}%
  \BibitemOpen
  \bibfield  {author} {\bibinfo {author} {\bibfnamefont {L.}~\bibnamefont
  {Benet}}, \bibinfo {author} {\bibfnamefont {F.}~\bibnamefont {Borgonovi}},
  \bibinfo {author} {\bibfnamefont {F.~M.}\ \bibnamefont {Izrailev}},\ and\
  \bibinfo {author} {\bibfnamefont {L.~F.}\ \bibnamefont {Santos}},\
  }\href@noop {} {\  (\bibinfo {year} {2022})},\ \Eprint
  {https://arxiv.org/abs/2211.10451} {arXiv:2211.10451 [cond-mat.stat-mech]}
  \BibitemShut {NoStop}%
\bibitem [{\citenamefont {Pilatowsky-Cameo}\ \emph {et~al.}(2020)\citenamefont
  {Pilatowsky-Cameo}, \citenamefont {Ch\'avez-Carlos}, \citenamefont
  {Bastarrachea-Magnani}, \citenamefont {Str\'ansk\'y}, \citenamefont
  {Lerma-Hern\'andez}, \citenamefont {Santos},\ and\ \citenamefont
  {Hirsch}}]{Pilatowsky2020OTOCregularSystem}%
  \BibitemOpen
  \bibfield  {author} {\bibinfo {author} {\bibfnamefont {S.}~\bibnamefont
  {Pilatowsky-Cameo}}, \bibinfo {author} {\bibfnamefont {J.}~\bibnamefont
  {Ch\'avez-Carlos}}, \bibinfo {author} {\bibfnamefont {M.~A.}\ \bibnamefont
  {Bastarrachea-Magnani}}, \bibinfo {author} {\bibfnamefont {P.}~\bibnamefont
  {Str\'ansk\'y}}, \bibinfo {author} {\bibfnamefont {S.}~\bibnamefont
  {Lerma-Hern\'andez}}, \bibinfo {author} {\bibfnamefont {L.~F.}\ \bibnamefont
  {Santos}},\ and\ \bibinfo {author} {\bibfnamefont {J.~G.}\ \bibnamefont
  {Hirsch}},\ }\href {https://doi.org/10.1103/PhysRevE.101.010202} {\bibfield
  {journal} {\bibinfo  {journal} {Phys. Rev. E}\ }\textbf {\bibinfo {volume}
  {101}},\ \bibinfo {pages} {010202} (\bibinfo {year} {2020})}\BibitemShut
  {NoStop}%
\bibitem [{\citenamefont {Kidd}, \citenamefont {Safavi-Naini},\ and\
  \citenamefont {Corney}(2021)}]{Kidd2021Dicke}%
  \BibitemOpen
  \bibfield  {author} {\bibinfo {author} {\bibfnamefont {R.~A.}\ \bibnamefont
  {Kidd}}, \bibinfo {author} {\bibfnamefont {A.}~\bibnamefont {Safavi-Naini}},\
  and\ \bibinfo {author} {\bibfnamefont {J.~F.}\ \bibnamefont {Corney}},\
  }\href {https://doi.org/10.1103/PhysRevA.103.033304} {\bibfield  {journal}
  {\bibinfo  {journal} {Phys. Rev. A}\ }\textbf {\bibinfo {volume} {103}},\
  \bibinfo {pages} {033304} (\bibinfo {year} {2021})}\BibitemShut {NoStop}%
\bibitem [{\citenamefont {Hashimoto}\ \emph {et~al.}(2020)\citenamefont
  {Hashimoto}, \citenamefont {Huh}, \citenamefont {Kim},\ and\ \citenamefont
  {Watanabe}}]{jhep11(2020)068}%
  \BibitemOpen
  \bibfield  {author} {\bibinfo {author} {\bibfnamefont {K.}~\bibnamefont
  {Hashimoto}}, \bibinfo {author} {\bibfnamefont {K.-B.}\ \bibnamefont {Huh}},
  \bibinfo {author} {\bibfnamefont {K.-Y.}\ \bibnamefont {Kim}},\ and\ \bibinfo
  {author} {\bibfnamefont {R.}~\bibnamefont {Watanabe}},\ }\href
  {https://doi.org/10.1007/jhep11(2020)068} {\bibfield  {journal} {\bibinfo
  {journal} {Journal of High Energy Physics}\ }\textbf {\bibinfo {volume}
  {2020}} (\bibinfo {year} {2020}),\ 10.1007/jhep11(2020)068}\BibitemShut
  {NoStop}%
\bibitem [{Note2()}]{Note2}%
  \BibitemOpen
  \bibinfo {note} {{ Interestingly, under specific circumstances, paradigmatic
  examples of systems with small local Hilbert space do have a well defined
  classical regime. This is the case with, for example, spin $1/2$ chains with
  all-to-all interactions \cite {Papparlardi2018}.}}\BibitemShut {Stop}%
\bibitem [{\citenamefont {Garc\'{\i}a-Mata}\ \emph {et~al.}(2018)\citenamefont
  {Garc\'{\i}a-Mata}, \citenamefont {Saraceno}, \citenamefont {Jalabert},
  \citenamefont {Roncaglia},\ and\ \citenamefont {Wisniacki}}]{argentinians}%
  \BibitemOpen
  \bibfield  {author} {\bibinfo {author} {\bibfnamefont {I.}~\bibnamefont
  {Garc\'{\i}a-Mata}}, \bibinfo {author} {\bibfnamefont {M.}~\bibnamefont
  {Saraceno}}, \bibinfo {author} {\bibfnamefont {R.~A.}\ \bibnamefont
  {Jalabert}}, \bibinfo {author} {\bibfnamefont {A.~J.}\ \bibnamefont
  {Roncaglia}},\ and\ \bibinfo {author} {\bibfnamefont {D.~A.}\ \bibnamefont
  {Wisniacki}},\ }\href {https://doi.org/10.1103/PhysRevLett.121.210601}
  {\bibfield  {journal} {\bibinfo  {journal} {Phys. Rev. Lett.}\ }\textbf
  {\bibinfo {volume} {121}},\ \bibinfo {pages} {210601} (\bibinfo {year}
  {2018})}\BibitemShut {NoStop}%
\bibitem [{\citenamefont {Fortes}\ \emph {et~al.}(2019)\citenamefont {Fortes},
  \citenamefont {Garcia-Mata}, \citenamefont {Jalabert},\ and\ \citenamefont
  {Wisniacki}}]{Fortes}%
  \BibitemOpen
  \bibfield  {author} {\bibinfo {author} {\bibfnamefont {E.}~\bibnamefont
  {Fortes}}, \bibinfo {author} {\bibfnamefont {I.}~\bibnamefont {Garcia-Mata}},
  \bibinfo {author} {\bibfnamefont {R.}~\bibnamefont {Jalabert}},\ and\
  \bibinfo {author} {\bibfnamefont {D.}~\bibnamefont {Wisniacki}},\ }\href
  {https://doi.org/10.1103/PhysRevE.100.042201} {\bibfield  {journal} {\bibinfo
   {journal} {Physical Review E}\ }\textbf {\bibinfo {volume} {100}} (\bibinfo
  {year} {2019}),\ 10.1103/PhysRevE.100.042201}\BibitemShut {NoStop}%
\bibitem [{\citenamefont {Polkovnikov}\ \emph {et~al.}(2011)\citenamefont
  {Polkovnikov}, \citenamefont {Sengupta}, \citenamefont {Silva},\ and\
  \citenamefont {Vengalattore}}]{Polkovnikov2011}%
  \BibitemOpen
  \bibfield  {author} {\bibinfo {author} {\bibfnamefont {A.}~\bibnamefont
  {Polkovnikov}}, \bibinfo {author} {\bibfnamefont {K.}~\bibnamefont
  {Sengupta}}, \bibinfo {author} {\bibfnamefont {A.}~\bibnamefont {Silva}},\
  and\ \bibinfo {author} {\bibfnamefont {M.}~\bibnamefont {Vengalattore}},\
  }\href {https://doi.org/10.1103/RevModPhys.83.863} {\bibfield  {journal}
  {\bibinfo  {journal} {Rev. Mod. Phys.}\ }\textbf {\bibinfo {volume} {83}},\
  \bibinfo {pages} {863} (\bibinfo {year} {2011})}\BibitemShut {NoStop}%
\bibitem [{\citenamefont {Schleich}(2001)}]{schleich01}%
  \BibitemOpen
  \bibfield  {author} {\bibinfo {author} {\bibfnamefont {W.~P.}\ \bibnamefont
  {Schleich}},\ }\href {https://doi.org/10.1002/3527602976} {\emph {\bibinfo
  {title} {Quantum Optics in Phase Space}}}\ (\bibinfo  {publisher}
  {Wiley-VCH},\ \bibinfo {address} {Berlin},\ \bibinfo {year}
  {2001})\BibitemShut {NoStop}%
\bibitem [{\citenamefont {Kim}\ and\ \citenamefont {Noz}(1991)}]{Kim1991}%
  \BibitemOpen
  \bibfield  {author} {\bibinfo {author} {\bibfnamefont {Y.~S.}\ \bibnamefont
  {Kim}}\ and\ \bibinfo {author} {\bibfnamefont {M.~E.}\ \bibnamefont {Noz}},\
  }\href {https://doi.org/10.1142/1197} {\emph {\bibinfo {title} {Phase Space
  Picture of Quantum Mechanics}}}\ (\bibinfo  {publisher} {World Scientific},\
  \bibinfo {year} {1991})\BibitemShut {NoStop}%
\bibitem [{\citenamefont {O'Connell}(2008)}]{connell2008}%
  \BibitemOpen
  \bibfield  {author} {\bibinfo {author} {\bibfnamefont {R.~F.}\ \bibnamefont
  {O'Connell}},\ }\href {https://doi.org/10.1142/S0219749908003451} {\bibfield
  {journal} {\bibinfo  {journal} {International Journal of Quantum
  Information}\ }\textbf {\bibinfo {volume} {06}},\ \bibinfo {pages} {415}
  (\bibinfo {year} {2008})}\BibitemShut {NoStop}%
\bibitem [{\citenamefont {Gutzwiller}(1991)}]{Gutzwiller1991}%
  \BibitemOpen
  \bibfield  {author} {\bibinfo {author} {\bibfnamefont {M.}~\bibnamefont
  {Gutzwiller}},\ }\href {https://doi.org/10.1007/978-1-4612-0983-6} {\emph
  {\bibinfo {title} {Chaos in Classical and Quantum Mechanics}}},\
  Interdisciplinary Applied Mathematics\ (\bibinfo  {publisher} {Springer New
  York},\ \bibinfo {year} {1991})\BibitemShut {NoStop}%
\bibitem [{\citenamefont {Haake}(2001)}]{Haake2001}%
  \BibitemOpen
  \bibfield  {author} {\bibinfo {author} {\bibfnamefont {F.}~\bibnamefont
  {Haake}},\ }\href {https://doi.org/10.1007/978-3-642-05428-0} {\emph
  {\bibinfo {title} {Quantum Signatures of Chaos}}},\ Physics and astronomy
  online library\ (\bibinfo  {publisher} {Springer},\ \bibinfo {year}
  {2001})\BibitemShut {NoStop}%
\bibitem [{\citenamefont {Brack}\ and\ \citenamefont
  {Bhaduri}(1997)}]{brack1997}%
  \BibitemOpen
  \bibfield  {author} {\bibinfo {author} {\bibfnamefont {M.}~\bibnamefont
  {Brack}}\ and\ \bibinfo {author} {\bibfnamefont {R.}~\bibnamefont
  {Bhaduri}},\ }\href {https://books.google.de/books?id=9mUsAAAAYAAJ} {\emph
  {\bibinfo {title} {Semiclassical Physics}}},\ Frontiers in physics\ (\bibinfo
   {publisher} {Avalon Publishing},\ \bibinfo {year} {1997})\BibitemShut
  {NoStop}%
\bibitem [{\citenamefont {Akila}\ \emph {et~al.}(2017)\citenamefont {Akila},
  \citenamefont {Waltner}, \citenamefont {Gutkin}, \citenamefont {Braun},\ and\
  \citenamefont {Guhr}}]{Waltner2017PRL}%
  \BibitemOpen
  \bibfield  {author} {\bibinfo {author} {\bibfnamefont {M.}~\bibnamefont
  {Akila}}, \bibinfo {author} {\bibfnamefont {D.}~\bibnamefont {Waltner}},
  \bibinfo {author} {\bibfnamefont {B.}~\bibnamefont {Gutkin}}, \bibinfo
  {author} {\bibfnamefont {P.}~\bibnamefont {Braun}},\ and\ \bibinfo {author}
  {\bibfnamefont {T.}~\bibnamefont {Guhr}},\ }\href
  {https://doi.org/10.1103/PhysRevLett.118.164101} {\bibfield  {journal}
  {\bibinfo  {journal} {Phys. Rev. Lett.}\ }\textbf {\bibinfo {volume} {118}},\
  \bibinfo {pages} {164101} (\bibinfo {year} {2017})}\BibitemShut {NoStop}%
\bibitem [{\citenamefont {Braun}\ \emph {et~al.}(2020)\citenamefont {Braun},
  \citenamefont {Waltner}, \citenamefont {Akila}, \citenamefont {Gutkin},\ and\
  \citenamefont {Guhr}}]{Waltner2020PRE}%
  \BibitemOpen
  \bibfield  {author} {\bibinfo {author} {\bibfnamefont {P.}~\bibnamefont
  {Braun}}, \bibinfo {author} {\bibfnamefont {D.}~\bibnamefont {Waltner}},
  \bibinfo {author} {\bibfnamefont {M.}~\bibnamefont {Akila}}, \bibinfo
  {author} {\bibfnamefont {B.}~\bibnamefont {Gutkin}},\ and\ \bibinfo {author}
  {\bibfnamefont {T.}~\bibnamefont {Guhr}},\ }\href
  {https://doi.org/10.1103/PhysRevE.101.052201} {\bibfield  {journal} {\bibinfo
   {journal} {Phys. Rev. E}\ }\textbf {\bibinfo {volume} {101}},\ \bibinfo
  {pages} {052201} (\bibinfo {year} {2020})}\BibitemShut {NoStop}%
\bibitem [{\citenamefont {Engl}, \citenamefont {Urbina},\ and\ \citenamefont
  {Richter}(2015)}]{Engl2015PRE}%
  \BibitemOpen
  \bibfield  {author} {\bibinfo {author} {\bibfnamefont {T.}~\bibnamefont
  {Engl}}, \bibinfo {author} {\bibfnamefont {J.~D.}\ \bibnamefont {Urbina}},\
  and\ \bibinfo {author} {\bibfnamefont {K.}~\bibnamefont {Richter}},\ }\href
  {https://doi.org/10.1103/PhysRevE.92.062907} {\bibfield  {journal} {\bibinfo
  {journal} {Phys. Rev. E}\ }\textbf {\bibinfo {volume} {92}},\ \bibinfo
  {pages} {062907} (\bibinfo {year} {2015})}\BibitemShut {NoStop}%
\bibitem [{\citenamefont {Tabor}(1989)}]{tabor1989chaos}%
  \BibitemOpen
  \bibfield  {author} {\bibinfo {author} {\bibfnamefont {M.}~\bibnamefont
  {Tabor}},\ }\href@noop {} {\emph {\bibinfo {title} {Chaos and Integrability
  in Nonlinear Dynamics: An Introduction}}}\ (\bibinfo  {publisher} {Wiley},\
  \bibinfo {year} {1989})\BibitemShut {NoStop}%
\bibitem [{\citenamefont {Wiggins}(2003)}]{wiggins2003}%
  \BibitemOpen
  \bibfield  {author} {\bibinfo {author} {\bibfnamefont {S.}~\bibnamefont
  {Wiggins}},\ }\href {https://doi.org/10.1007/b97481} {\emph {\bibinfo {title}
  {Introduction to Applied Nonlinear Dynamical Systems and Chaos}}},\ Texts in
  Applied Mathematics\ (\bibinfo  {publisher} {Springer New York},\ \bibinfo
  {year} {2003})\BibitemShut {NoStop}%
\bibitem [{\citenamefont {Est{\`{e}}ve}\ \emph {et~al.}(2008)\citenamefont
  {Est{\`{e}}ve}, \citenamefont {Gross}, \citenamefont {Weller}, \citenamefont
  {Giovanazzi},\ and\ \citenamefont {Oberthaler}}]{squeezingBEC2008}%
  \BibitemOpen
  \bibfield  {author} {\bibinfo {author} {\bibfnamefont {J.}~\bibnamefont
  {Est{\`{e}}ve}}, \bibinfo {author} {\bibfnamefont {C.}~\bibnamefont {Gross}},
  \bibinfo {author} {\bibfnamefont {A.}~\bibnamefont {Weller}}, \bibinfo
  {author} {\bibfnamefont {S.}~\bibnamefont {Giovanazzi}},\ and\ \bibinfo
  {author} {\bibfnamefont {M.~K.}\ \bibnamefont {Oberthaler}},\ }\href
  {https://doi.org/10.1038/nature07332} {\bibfield  {journal} {\bibinfo
  {journal} {Nature}\ }\textbf {\bibinfo {volume} {455}},\ \bibinfo {pages}
  {1216} (\bibinfo {year} {2008})}\BibitemShut {NoStop}%
\bibitem [{\citenamefont {Jalabert}, \citenamefont {Garc\'{\i}a-Mata},\ and\
  \citenamefont {Wisniacki}(2018)}]{PhysRevE.98.062218}%
  \BibitemOpen
  \bibfield  {author} {\bibinfo {author} {\bibfnamefont {R.~A.}\ \bibnamefont
  {Jalabert}}, \bibinfo {author} {\bibfnamefont {I.}~\bibnamefont
  {Garc\'{\i}a-Mata}},\ and\ \bibinfo {author} {\bibfnamefont {D.~A.}\
  \bibnamefont {Wisniacki}},\ }\href
  {https://doi.org/10.1103/PhysRevE.98.062218} {\bibfield  {journal} {\bibinfo
  {journal} {Phys. Rev. E}\ }\textbf {\bibinfo {volume} {98}},\ \bibinfo
  {pages} {062218} (\bibinfo {year} {2018})}\BibitemShut {NoStop}%
\bibitem [{\citenamefont {Geiger}(2020)}]{thesisBenni}%
  \BibitemOpen
  \bibfield  {author} {\bibinfo {author} {\bibfnamefont {B.}~\bibnamefont
  {Geiger}},\ }\href {https://epub.uni-regensburg.de/43571/} {\emph {\bibinfo
  {title} {From few to many particles: Semiclassical approaches to interacting
  quantum systems}}},\ Vol.~\bibinfo {volume} {55}\ (\bibinfo  {publisher}
  {University of Regensburg},\ \bibinfo {year} {2020})\BibitemShut {NoStop}%
\bibitem [{\citenamefont {Il'ichev}\ \emph {et~al.}(2001)\citenamefont
  {Il'ichev}, \citenamefont {Grajcar}, \citenamefont {Hlubina}, \citenamefont
  {IJsselsteijn}, \citenamefont {Hoenig}, \citenamefont {Meyer}, \citenamefont
  {Golubov}, \citenamefont {Amin}, \citenamefont {Zagoskin}, \citenamefont
  {Omelyanchouk},\ and\ \citenamefont
  {Kupriyanov}}]{superconductorJosephson2001}%
  \BibitemOpen
  \bibfield  {author} {\bibinfo {author} {\bibfnamefont {E.}~\bibnamefont
  {Il'ichev}}, \bibinfo {author} {\bibfnamefont {M.}~\bibnamefont {Grajcar}},
  \bibinfo {author} {\bibfnamefont {R.}~\bibnamefont {Hlubina}}, \bibinfo
  {author} {\bibfnamefont {R.~P.~J.}\ \bibnamefont {IJsselsteijn}}, \bibinfo
  {author} {\bibfnamefont {H.~E.}\ \bibnamefont {Hoenig}}, \bibinfo {author}
  {\bibfnamefont {H.-G.}\ \bibnamefont {Meyer}}, \bibinfo {author}
  {\bibfnamefont {A.}~\bibnamefont {Golubov}}, \bibinfo {author} {\bibfnamefont
  {M.~H.~S.}\ \bibnamefont {Amin}}, \bibinfo {author} {\bibfnamefont {A.~M.}\
  \bibnamefont {Zagoskin}}, \bibinfo {author} {\bibfnamefont {A.~N.}\
  \bibnamefont {Omelyanchouk}},\ and\ \bibinfo {author} {\bibfnamefont {M.~Y.}\
  \bibnamefont {Kupriyanov}},\ }\href
  {https://doi.org/10.1103/PhysRevLett.86.5369} {\bibfield  {journal} {\bibinfo
   {journal} {Phys. Rev. Lett.}\ }\textbf {\bibinfo {volume} {86}},\ \bibinfo
  {pages} {5369} (\bibinfo {year} {2001})}\BibitemShut {NoStop}%
\bibitem [{\citenamefont {Albiez}\ \emph {et~al.}(2005)\citenamefont {Albiez},
  \citenamefont {Gati}, \citenamefont {F\"olling}, \citenamefont {Hunsmann},
  \citenamefont {Cristiani},\ and\ \citenamefont {Oberthaler}}]{Albiez2005}%
  \BibitemOpen
  \bibfield  {author} {\bibinfo {author} {\bibfnamefont {M.}~\bibnamefont
  {Albiez}}, \bibinfo {author} {\bibfnamefont {R.}~\bibnamefont {Gati}},
  \bibinfo {author} {\bibfnamefont {J.}~\bibnamefont {F\"olling}}, \bibinfo
  {author} {\bibfnamefont {S.}~\bibnamefont {Hunsmann}}, \bibinfo {author}
  {\bibfnamefont {M.}~\bibnamefont {Cristiani}},\ and\ \bibinfo {author}
  {\bibfnamefont {M.~K.}\ \bibnamefont {Oberthaler}},\ }\href
  {https://doi.org/10.1103/PhysRevLett.95.010402} {\bibfield  {journal}
  {\bibinfo  {journal} {Phys. Rev. Lett.}\ }\textbf {\bibinfo {volume} {95}},\
  \bibinfo {pages} {010402} (\bibinfo {year} {2005})}\BibitemShut {NoStop}%
\bibitem [{\citenamefont {Fölling}\ \emph {et~al.}(2007)\citenamefont
  {Fölling}, \citenamefont {Trotzky}, \citenamefont {Cheinet}, \citenamefont
  {Feld}, \citenamefont {Saers}, \citenamefont {Widera}, \citenamefont
  {Müller},\ and\ \citenamefont {Bloch}}]{F_lling_2007}%
  \BibitemOpen
  \bibfield  {author} {\bibinfo {author} {\bibfnamefont {S.}~\bibnamefont
  {Fölling}}, \bibinfo {author} {\bibfnamefont {S.}~\bibnamefont {Trotzky}},
  \bibinfo {author} {\bibfnamefont {P.}~\bibnamefont {Cheinet}}, \bibinfo
  {author} {\bibfnamefont {M.}~\bibnamefont {Feld}}, \bibinfo {author}
  {\bibfnamefont {R.}~\bibnamefont {Saers}}, \bibinfo {author} {\bibfnamefont
  {A.}~\bibnamefont {Widera}}, \bibinfo {author} {\bibfnamefont
  {T.}~\bibnamefont {Müller}},\ and\ \bibinfo {author} {\bibfnamefont
  {I.}~\bibnamefont {Bloch}},\ }\href {https://doi.org/10.1038/nature06112}
  {\bibfield  {journal} {\bibinfo  {journal} {Nature}\ }\textbf {\bibinfo
  {volume} {448}},\ \bibinfo {pages} {1029} (\bibinfo {year}
  {2007})}\BibitemShut {NoStop}%
\bibitem [{\citenamefont {Witthaut}, \citenamefont {Trimborn},\ and\
  \citenamefont {Wimberger}(2008)}]{Witthaut2008}%
  \BibitemOpen
  \bibfield  {author} {\bibinfo {author} {\bibfnamefont {D.}~\bibnamefont
  {Witthaut}}, \bibinfo {author} {\bibfnamefont {F.}~\bibnamefont {Trimborn}},\
  and\ \bibinfo {author} {\bibfnamefont {S.}~\bibnamefont {Wimberger}},\ }\href
  {https://doi.org/10.1103/PhysRevLett.101.200402} {\bibfield  {journal}
  {\bibinfo  {journal} {Phys. Rev. Lett.}\ }\textbf {\bibinfo {volume} {101}},\
  \bibinfo {pages} {200402} (\bibinfo {year} {2008})}\BibitemShut {NoStop}%
\bibitem [{\citenamefont {Cheinet}\ \emph {et~al.}(2008)\citenamefont
  {Cheinet}, \citenamefont {Trotzky}, \citenamefont {Feld}, \citenamefont
  {Schnorrberger}, \citenamefont {Moreno-Cardoner}, \citenamefont {F\"olling},\
  and\ \citenamefont {Bloch}}]{cheinet2008}%
  \BibitemOpen
  \bibfield  {author} {\bibinfo {author} {\bibfnamefont {P.}~\bibnamefont
  {Cheinet}}, \bibinfo {author} {\bibfnamefont {S.}~\bibnamefont {Trotzky}},
  \bibinfo {author} {\bibfnamefont {M.}~\bibnamefont {Feld}}, \bibinfo {author}
  {\bibfnamefont {U.}~\bibnamefont {Schnorrberger}}, \bibinfo {author}
  {\bibfnamefont {M.}~\bibnamefont {Moreno-Cardoner}}, \bibinfo {author}
  {\bibfnamefont {S.}~\bibnamefont {F\"olling}},\ and\ \bibinfo {author}
  {\bibfnamefont {I.}~\bibnamefont {Bloch}},\ }\href
  {https://doi.org/10.1103/PhysRevLett.101.090404} {\bibfield  {journal}
  {\bibinfo  {journal} {Phys. Rev. Lett.}\ }\textbf {\bibinfo {volume} {101}},\
  \bibinfo {pages} {090404} (\bibinfo {year} {2008})}\BibitemShut {NoStop}%
\bibitem [{\citenamefont {Kierig}\ \emph {et~al.}(2008)\citenamefont {Kierig},
  \citenamefont {Schnorrberger}, \citenamefont {Schietinger}, \citenamefont
  {Tomkovic},\ and\ \citenamefont {Oberthaler}}]{Kierig2008}%
  \BibitemOpen
  \bibfield  {author} {\bibinfo {author} {\bibfnamefont {E.}~\bibnamefont
  {Kierig}}, \bibinfo {author} {\bibfnamefont {U.}~\bibnamefont
  {Schnorrberger}}, \bibinfo {author} {\bibfnamefont {A.}~\bibnamefont
  {Schietinger}}, \bibinfo {author} {\bibfnamefont {J.}~\bibnamefont
  {Tomkovic}},\ and\ \bibinfo {author} {\bibfnamefont {M.~K.}\ \bibnamefont
  {Oberthaler}},\ }\href {https://doi.org/10.1103/PhysRevLett.100.190405}
  {\bibfield  {journal} {\bibinfo  {journal} {Phys. Rev. Lett.}\ }\textbf
  {\bibinfo {volume} {100}},\ \bibinfo {pages} {190405} (\bibinfo {year}
  {2008})}\BibitemShut {NoStop}%
\bibitem [{\citenamefont {Tomkovi\ifmmode~\check{c}\else \v{c}\fi{}}\ \emph
  {et~al.}(2017)\citenamefont {Tomkovi\ifmmode~\check{c}\else \v{c}\fi{}},
  \citenamefont {Muessel}, \citenamefont {Strobel}, \citenamefont {L\"ock},
  \citenamefont {Schlagheck}, \citenamefont {Ketzmerick},\ and\ \citenamefont
  {Oberthaler}}]{Tomkovi2017}%
  \BibitemOpen
  \bibfield  {author} {\bibinfo {author} {\bibfnamefont {J.}~\bibnamefont
  {Tomkovi\ifmmode~\check{c}\else \v{c}\fi{}}}, \bibinfo {author}
  {\bibfnamefont {W.}~\bibnamefont {Muessel}}, \bibinfo {author} {\bibfnamefont
  {H.}~\bibnamefont {Strobel}}, \bibinfo {author} {\bibfnamefont
  {S.}~\bibnamefont {L\"ock}}, \bibinfo {author} {\bibfnamefont
  {P.}~\bibnamefont {Schlagheck}}, \bibinfo {author} {\bibfnamefont
  {R.}~\bibnamefont {Ketzmerick}},\ and\ \bibinfo {author} {\bibfnamefont
  {M.~K.}\ \bibnamefont {Oberthaler}},\ }\href
  {https://doi.org/10.1103/PhysRevA.95.011602} {\bibfield  {journal} {\bibinfo
  {journal} {Phys. Rev. A}\ }\textbf {\bibinfo {volume} {95}},\ \bibinfo
  {pages} {011602} (\bibinfo {year} {2017})}\BibitemShut {NoStop}%
\bibitem [{\citenamefont {Negele}\ and\ \citenamefont
  {Orland}(1995)}]{negele1995quantum}%
  \BibitemOpen
  \bibfield  {author} {\bibinfo {author} {\bibfnamefont {J.}~\bibnamefont
  {Negele}}\ and\ \bibinfo {author} {\bibfnamefont {H.}~\bibnamefont
  {Orland}},\ }\href {https://books.google.fr/books?id=80hkPgAACAAJ} {\emph
  {\bibinfo {title} {Quantum Many Particle Systems}}}\ (\bibinfo  {publisher}
  {Basic Books},\ \bibinfo {year} {1995})\BibitemShut {NoStop}%
\bibitem [{\citenamefont {Campbell}(2020)}]{Campbell2020Dimer}%
  \BibitemOpen
  \bibfield  {author} {\bibinfo {author} {\bibfnamefont {D.~K.}\ \bibnamefont
  {Campbell}},\ }in\ \href {https://doi.org/10.1007/978-3-030-35473-2_9} {\emph
  {\bibinfo {booktitle} {Strongly Coupled Field Theories for Condensed Matter
  and Quantum Information Theory}}},\ \bibinfo {editor} {edited by\ \bibinfo
  {editor} {\bibfnamefont {A.}~\bibnamefont {Ferraz}}, \bibinfo {editor}
  {\bibfnamefont {K.~S.}\ \bibnamefont {Gupta}}, \bibinfo {editor}
  {\bibfnamefont {G.~W.}\ \bibnamefont {Semenoff}},\ and\ \bibinfo {editor}
  {\bibfnamefont {P.}~\bibnamefont {Sodano}}}\ (\bibinfo  {publisher} {Springer
  International Publishing},\ \bibinfo {address} {Cham},\ \bibinfo {year}
  {2020})\ pp.\ \bibinfo {pages} {247--258}\BibitemShut {NoStop}%
\bibitem [{\citenamefont {Gardiner}, \citenamefont {Zoller},\ and\
  \citenamefont {Zoller}(2004)}]{gardiner2004quantum}%
  \BibitemOpen
  \bibfield  {author} {\bibinfo {author} {\bibfnamefont {C.}~\bibnamefont
  {Gardiner}}, \bibinfo {author} {\bibfnamefont {P.}~\bibnamefont {Zoller}},\
  and\ \bibinfo {author} {\bibfnamefont {P.}~\bibnamefont {Zoller}},\ }\href
  {https://books.google.de/books?id=a\_xsT8oGhdgC} {\emph {\bibinfo {title}
  {Quantum Noise: A Handbook of Markovian and Non-Markovian Quantum Stochastic
  Methods with Applications to Quantum Optics}}},\ Springer Series in
  Synergetics\ (\bibinfo  {publisher} {Springer},\ \bibinfo {year}
  {2004})\BibitemShut {NoStop}%
\bibitem [{\citenamefont {Lignier}\ \emph {et~al.}(2007)\citenamefont
  {Lignier}, \citenamefont {Sias}, \citenamefont {Ciampini}, \citenamefont
  {Singh}, \citenamefont {Zenesini}, \citenamefont {Morsch},\ and\
  \citenamefont {Arimondo}}]{Lignier2007}%
  \BibitemOpen
  \bibfield  {author} {\bibinfo {author} {\bibfnamefont {H.}~\bibnamefont
  {Lignier}}, \bibinfo {author} {\bibfnamefont {C.}~\bibnamefont {Sias}},
  \bibinfo {author} {\bibfnamefont {D.}~\bibnamefont {Ciampini}}, \bibinfo
  {author} {\bibfnamefont {Y.}~\bibnamefont {Singh}}, \bibinfo {author}
  {\bibfnamefont {A.}~\bibnamefont {Zenesini}}, \bibinfo {author}
  {\bibfnamefont {O.}~\bibnamefont {Morsch}},\ and\ \bibinfo {author}
  {\bibfnamefont {E.}~\bibnamefont {Arimondo}},\ }\href
  {https://doi.org/10.1103/PhysRevLett.99.220403} {\bibfield  {journal}
  {\bibinfo  {journal} {Phys. Rev. Lett.}\ }\textbf {\bibinfo {volume} {99}},\
  \bibinfo {pages} {220403} (\bibinfo {year} {2007})}\BibitemShut {NoStop}%
\bibitem [{\citenamefont {Rozenbaum}, \citenamefont {Ganeshan},\ and\
  \citenamefont {Galitski}(2017)}]{Rozenbaum_2017}%
  \BibitemOpen
  \bibfield  {author} {\bibinfo {author} {\bibfnamefont {E.~B.}\ \bibnamefont
  {Rozenbaum}}, \bibinfo {author} {\bibfnamefont {S.}~\bibnamefont
  {Ganeshan}},\ and\ \bibinfo {author} {\bibfnamefont {V.}~\bibnamefont
  {Galitski}},\ }\href {https://doi.org/10.1103/physrevlett.118.086801}
  {\bibfield  {journal} {\bibinfo  {journal} {Physical Review Letters}\
  }\textbf {\bibinfo {volume} {118}} (\bibinfo {year} {2017}),\
  10.1103/physrevlett.118.086801}\BibitemShut {NoStop}%
\bibitem [{\citenamefont {Lakshminarayan}(2019)}]{arul2019}%
  \BibitemOpen
  \bibfield  {author} {\bibinfo {author} {\bibfnamefont {A.}~\bibnamefont
  {Lakshminarayan}},\ }\href {https://doi.org/10.1103/PhysRevE.99.012201}
  {\bibfield  {journal} {\bibinfo  {journal} {Phys. Rev. E}\ }\textbf {\bibinfo
  {volume} {99}},\ \bibinfo {pages} {012201} (\bibinfo {year}
  {2019})}\BibitemShut {NoStop}%
\bibitem [{\citenamefont {Meier}\ \emph {et~al.}(2023)\citenamefont {Meier},
  \citenamefont {Steinhuber}, \citenamefont {Urbina}, \citenamefont {Waltner},\
  and\ \citenamefont {Guhr}}]{meier2023signatures}%
  \BibitemOpen
  \bibfield  {author} {\bibinfo {author} {\bibfnamefont {F.}~\bibnamefont
  {Meier}}, \bibinfo {author} {\bibfnamefont {M.}~\bibnamefont {Steinhuber}},
  \bibinfo {author} {\bibfnamefont {J.~D.}\ \bibnamefont {Urbina}}, \bibinfo
  {author} {\bibfnamefont {D.}~\bibnamefont {Waltner}},\ and\ \bibinfo {author}
  {\bibfnamefont {T.}~\bibnamefont {Guhr}},\ }\href
  {https://doi.org/10.1103/PhysRevE.107.054202} {\bibfield  {journal} {\bibinfo
   {journal} {Phys. Rev. E}\ }\textbf {\bibinfo {volume} {107}},\ \bibinfo
  {pages} {054202} (\bibinfo {year} {2023})}\BibitemShut {NoStop}%
\bibitem [{\citenamefont {Case}(2008)}]{Case2008}%
  \BibitemOpen
  \bibfield  {author} {\bibinfo {author} {\bibfnamefont {W.~B.}\ \bibnamefont
  {Case}},\ }\href {https://doi.org/10.1119/1.2957889} {\bibfield  {journal}
  {\bibinfo  {journal} {American Journal of Physics}\ }\textbf {\bibinfo
  {volume} {76}},\ \bibinfo {pages} {937} (\bibinfo {year} {2008})}\BibitemShut
  {NoStop}%
\bibitem [{\citenamefont {Curtright}\ and\ \citenamefont
  {Zachos}(2012)}]{Curtright2012}%
  \BibitemOpen
  \bibfield  {author} {\bibinfo {author} {\bibfnamefont {T.~L.}\ \bibnamefont
  {Curtright}}\ and\ \bibinfo {author} {\bibfnamefont {C.~K.}\ \bibnamefont
  {Zachos}},\ }\href {https://doi.org/10.1142/S2251158X12000069} {\bibfield
  {journal} {\bibinfo  {journal} {Asia Pacific Physics Newsletter}\ }\textbf
  {\bibinfo {volume} {01}},\ \bibinfo {pages} {37} (\bibinfo {year}
  {2012})}\BibitemShut {NoStop}%
\end{thebibliography}%
\end{document}